# Line Transversals of Convex Polyhedra in $\mathbb{R}^3$[*]

Haim Kaplan[†]    Natan Rubin[‡]    Micha Sharir[§]

October 26, 2018

## Abstract

We establish a bound of $O(n^2 k^{1+\varepsilon})$, for any $\varepsilon > 0$, on the combinatorial complexity of the set $\mathcal{T}$ of line transversals of a collection $\mathcal{P}$ of $k$ convex polyhedra in $\mathbb{R}^3$ with a total of $n$ facets, and present a randomized algorithm which computes the boundary of $\mathcal{T}$ in comparable expected time. Thus, when $k \ll n$, the new bounds on the complexity (and construction cost) of $\mathcal{T}$ improve upon the previously best known bounds, which are nearly cubic in $n$.

To obtain the above result, we study the set $\mathcal{T}_{\ell_0}$ of line transversals which emanate from a fixed line $\ell_0$, establish an almost tight bound of $O(nk^{1+\varepsilon})$ on the complexity of $\mathcal{T}_{\ell_0}$, and provide a randomized algorithm which computes $\mathcal{T}_{\ell_0}$ in comparable expected time. Slightly improved combinatorial bounds for the complexity of $\mathcal{T}_{\ell_0}$, and comparable improvements in the cost of constructing this set, are established for two special cases, both assuming that the polyhedra of $\mathcal{P}$ are pairwise disjoint: the case where $\ell_0$ is disjoint from the polyhedra of $\mathcal{P}$, and the case where the polyhedra of $\mathcal{P}$ are unbounded in a direction parallel to $\ell_0$.

[*]Work by Haim Kaplan and Natan Rubin has been supported by Grant 975/06 from the Israel Science Foundation and by a the United states - Israel Binational Science Foundation, grant number 2006204. Work by Micha Sharir was partially supported by NSF Grant CCF-05-14079, by a grant from the U.S.-Israeli Binational Science Foundation, by grant 155/05 from the Israel Science Fund, Israeli Academy of Sciences, by a grant from the AFIRST joint French-Israeli program, and by the Hermann Minkowski–MINERVA Center for Geometry at Tel Aviv University.

[†]School of Computer Science, Tel Aviv University, Tel Aviv 69978, Israel. E-mail: `haimk@post.tau.ac.il`
[‡]School of Computer Science, Tel Aviv University, Tel Aviv 69978, Israel. E-mail: `rubinnat@post.tau.ac.il`
[§]School of Computer Science, Tel Aviv University, Tel Aviv 69978, Israel, and Courant Institute of Mathematical Sciences, New York University, New York, NY 10012, USA. E-mail: `michas@post.tau.ac.il`

# 1 Introduction

**Line transversals—a brief background.** In this paper we study the combinatorial complexity of the set of line transversals of a collection $\mathcal{P}$ of $k$ convex polyhedra in $\mathbb{R}^3$ with a total of $n$ facets. This is a special case of the general study of line transversals to a collection of convex sets in $\mathbb{R}^d$, for any $d \geq 2$, a topic that has been extensively studied for several decades; see the survey papers [15, 25, 31].

Let $\mathcal{P}$ be a family of $k$ convex sets in $\mathbb{R}^3$. A line $\ell$ is a *transversal* of $\mathcal{P}$ if it intersects every member of $\mathcal{P}$. The set of all line transversals of $\mathcal{P}$ is called the *transversal space* (or *stabbing region* of $\mathcal{P}$), and is denoted by $\mathcal{T}(\mathcal{P})$. The *combinatorial complexity* of $\mathcal{T}(\mathcal{P})$ is defined as the total number of topological faces, of all dimensions, on the boundary of $\mathcal{T}(\mathcal{P})$. If the objects in $\mathcal{P}$ are pairwise disjoint, a line transversal meets them in a well-defined order, called a *geometric permutation*.

A standard reduction leads to a representation of $\mathcal{T}(\mathcal{P})$ as a region (a "sandwich" region) in $\mathbb{R}^4$, enclosed between the lower envelope and the upper envelope of two collections of surfaces, describing upper and lower tangencies to each member of $\mathcal{P}$; see for example [2, 21] for a description of this reduction, a special case of which is described later in this paper. When $\mathcal{P}$ is a set of convex polyhedra, the case considered in this paper, the surface of line tangents to any $P \in \mathcal{P}$ is decomposed into patches, each representing tangents to $P$ at a fixed edge of $P$. The boundary vertices of $\mathcal{T}(\mathcal{P})$ correspond to *extremal stabbing lines*, which are transversals of $\mathcal{P}$ that (i) are tangent to some polyhedra of $\mathcal{P}$, and (ii) cannot be continuously moved while continuing to touch the same edges and vertices of those polyhedra. As we will later note, the worst-case combinatorial complexity of $\mathcal{T}(\mathcal{P})$ is bounded by the maximum number of vertices of $\mathcal{T}(\mathcal{P})$, so it suffices to bound the latter quantity.

Any upper bound on the maximal *combinatorial complexity* of $\mathcal{T}(\mathcal{P})$ also serves as a natural upper bound on the maximal number of connected components of $\mathcal{T}(\mathcal{P})$ (and, aposteriori, also on the number of geometric permutations of $\mathcal{P}$, assuming that its elements are pairwise disjoint).

In the planar case, the complexity of $\mathcal{T}(\mathcal{P})$, when the elements of $\mathcal{P}$ are pairwise disjoint, is $O(k)$ (see, e.g., [12]), but can be $\Omega(n)$ otherwise, where $n$ is the total description complexity of the objects of $\mathcal{T}$, with a matching slightly super-linear upper bound (in $n$). (E.g., when the objects of $\mathcal{P}$ are convex polygons, and $n$ is the overall number of their edges, the complexity of $\mathcal{T}(\mathcal{P})$ is $O(n\alpha(n))$, where $\alpha(n)$ is the slowly-growing inverse Ackermann function.) In the 3-dimensional case, the complexity of $\mathcal{T}(\mathcal{P})$ depends on $n$, even if the elements of $\mathcal{P}$ are pairwise disjoint. The first algorithms for computing $\mathcal{T}(\mathcal{P})$, where $\mathcal{P}$ is a set of convex polyhedra, with a total of $n$ facets, run in time about $O(n^4)$ [6, 22]. Pellegrini and Shor [24] establish an upper bound of $O(n^{3+\varepsilon})$ on the complexity of $\mathcal{T}(\mathcal{P})$. They also describe an algorithm for computing the boundary vertices of $\mathcal{T}(\mathcal{P})$, with a comparable running time. Agarwal [1] improves the upper bound for the complexity of $\mathcal{T}(\mathcal{P})$ to $O(n^3 \log n)$. When the sets in $\mathcal{P}$ are semi-algebraic of *constant description complexity*[1] (for example, if the sets in $\mathcal{P}$ are balls, or tetrahedra), the complexity of $\mathcal{T}(\mathcal{P})$ is $O(n^{3+\varepsilon})$, for any $\varepsilon > 0$, as follows from the general and more recent result of Koltun and Sharir [21] on the complexity of sandwich regions of trivariate functions.[2] In $\mathbb{R}^3$ there are almost matching lower-bound constructions, showing that the complexity of $\mathcal{T}(\mathcal{P})$ can be $\Omega(n^3)$; see [2, 23]. Those lower bounds are constructed using collections of $n$ triangles in $\mathbb{R}^3$. However, for the number of connected components of $\mathcal{T}(\mathcal{P})$, the best known lower bounds are $\Omega(n^2)$, or $\Omega(n^{d-1})$ in $d$ dimensions [28].

---

[1]That is, each set is a semi-algebraic set defined by a Boolean combination of a constant number of polynomial equalities and inequalities of constant maximum degree.

[2]It is indeed stated in [21] as a corollary.

Narrowing this gap, even for restricted families of objects, is an intriguing open problem, already for $d = 3$ dimensions.

These are the best general known bounds on the complexity of $\mathcal{T}(\mathcal{P})$, but there are some improved bounds in restricted cases: Aronov and Smorodinsky [5] proved that when restricting the transversals to pass through a fixed point, the transversal space has a maximum of $\Theta(k^{d-1})$ components, for any collection $\mathcal{P}$ of $k$ (not necessarily pairwise disjoint) convex sets in $\mathbb{R}^d$. Brönnimann et al. [8] gave a complete description of the transversal space of $k$ segments in $\mathbb{R}^3$. In this case the transversal space consists of a maximum of $\Theta(k)$ connected components.

Another line of research initiated by Katchalski et al. [18] studies the number $g_d(k)$ of geometric permutations of a set $\mathcal{P}$ of pairwise disjoint convex objects. The known bounds on $g_d(k)$ in the general case are: $g_2(k) = 2k-2$; see [12], $g_d(k) = O(k^{2d-2}), d \geq 3$; see [30], and $g_d(k) = \Omega(k^{d-1}), d \geq 3$; see [19, 28]. Improved bounds are known in several special cases such as pairwise disjoint balls [28], and families of fat objects [20].

One of the most relevant predecessors of this paper is a recent paper of Brönnimann et al. [7], who prove that the entire arrangement $\mathcal{A}^*$ of surfaces describing upper and lower tangencies to the individual polyhedra of $\mathcal{P}$ has complexity $O(n^2 k^2)$, and this bound is tight in the worst case. Each cell in $\mathcal{A}^*$ corresponds to a maximal connected set of lines which stab the same subset of $\mathcal{P}$. In particular, $\mathcal{T}(\mathcal{P})$ is equal to the union of all the cells in $\mathcal{A}^*(\mathcal{P})$ whose stabbed subset is the entire $\mathcal{P}$.

Efrat et al. [13] assume that the polyhedra in $\mathcal{P}$ are pairwise disjoint, and consider the restricted space $\mathcal{L}$ of lines that pass through a fixed line $\ell_0$. They prove that the set of all lines in $\mathcal{L}$ which stab at least one polyhedron in $\mathcal{P}$ (alternatively, the set of all lines in $\mathcal{L}$ which miss all the polyhedra of $\mathcal{P}$) has combinatorial complexity $O(nk^2)$, and this bound is tight in the worst case. When the polyhedra in $\mathcal{P}$ are unbounded in the $\ell_0$-negative direction, the bound improves to $O(nk2^{\alpha(k)})$; see [13] for more details.

**Our contribution.** We consider an arbitrary collection $\mathcal{P}$ of $k$ convex polyhedra in $\mathbb{R}^3$, with a total of $n$ facets, and derive an upper bound of $O(n^2 k^{1+\varepsilon})$, for any $\varepsilon > 0$, on the complexity of $\mathcal{T}(\mathcal{P})$. We provide a randomized (Las-Vegas) algorithm for computing a description of the boundary of $\mathcal{T}(\mathcal{P})$, with comparable expected running time. We also present a lower bound construction of such a collection $\mathcal{P}$, for which $\mathcal{T}(\mathcal{P})$ has complexity $\Omega(n^2 + nk^2)$.

To achieve the general upper bound on the complexity of $\mathcal{T}(\mathcal{P})$, we focus on the restricted case where we only consider lines which pass through a fixed line $\ell_0$, and study the resulting stabbing region $\mathcal{T}_{\ell_0}(\mathcal{P}) := \mathcal{T}(\mathcal{P}) \cap \mathcal{L}$, where $\mathcal{L} = \mathcal{L}_{\ell_0}$ is the space of these restricted lines. Unlike the general case of lines in 3-space, which have four degrees of freedom, lines in $\mathcal{L}$ have only three degrees of freedom, and we represent them as points in an appropriately parametrized 3-dimensional space. The overall bound is obtained by repeating this analysis for all the $O(n)$ lines $\ell_0$ which contain polyhedra edges.

The *combinatorial complexity* of $\mathcal{T}_{\ell_0}(\mathcal{P})$ can be measured in terms of the number of its *vertices*, each representing an extremal stabbing line, which passes through (the polyhedron edge contained in) $\ell_0$. We show that the combinatorial complexity of $\mathcal{T}_{\ell_0}(\mathcal{P})$ is $O(nk^{1+\varepsilon})$, for any $\varepsilon > 0$, and that the boundary representation of $\mathcal{T}_{\ell_0}(\mathcal{P})$ can be computed in comparable randomized expected time. To appreciate this bound, we note that the standard representation of $\mathcal{T}(\mathcal{P})$ as a sandwich region between a lower envelope and an upper envelope also holds in the case of $\mathcal{T}_{\ell_0}(\mathcal{P})$, except that here the envelopes are of *bivariate* functions. There are only $k$ functions in each collection, where each function represents an upper tangency or a lower tangency to some fixed polyhedron in $\mathcal{P}$, but their graphs do not have constant description complexity—each graph is partitioned into patches, each representing tangency at some fixed edge of the respective polyhedron. We can thus regard the sandwich region as being formed by a total of $O(n)$ partially defined bivariate functions, each now of constant description complexity, so, by the results of [3, 21], the complexity of the stabbing region is $O(n^{2+\varepsilon})$, for any $\varepsilon > 0$. Our contribution is thus in making this bound depend also on $k$, so that it becomes only *linear* in $n$; this is a significant improvement when $k \ll n$.

We also consider a pair of restricted instances of the problem, both of which assume that the polyhedra of $\mathcal{P}$ are pairwise disjoint. In the first case, when the polyhedra of $\mathcal{P}$ are disjoint from $\ell_0$, our general analysis easily implies that the complexity of the stabbing region is only $O((nk + k^3)\beta_4(k))$, where $\beta_4(k) = 2^{\alpha(k)}$ is an extremely slowly growing function[3], expressed in terms of the inverse Ackermann function $\alpha(k)$. In the second case, when all the polyhedra in $\mathcal{P}$ are unbounded in a direction parallel to $\ell_0$, we improve the upper bound on the stabbing region to $O(nk\beta_4(k))$. In this case, the sandwich region degenerates to the region above the upper envelope of the lower tangency functions, or, symmetrically, to the region below the lower envelope of the upper tangency functions. The improved bound then follows by showing that the complexity of a single envelope, rather than of a sandwich region between two envelopes, is only $O(nk\beta_4(k))$. (This bound holds regardless of whether the polyhedra are unbounded or not, but it requires them to be pairwise disjoint.) In both special instances, the stabbing region within $\mathcal{L}$ can be computed in deterministic time, asymptotically close to its worst-case complexity.

We also show that the number of geometric permutations of $\mathcal{P}$ induced by lines in $\mathcal{L}$ is $O(\min\{k^3, nk^{1+\varepsilon}\})$. A naive bound on this number is $O(k^4)$ (which is Wenger's general bound mentioned above [30]). The second term in this bound follows from the fact that the number of geometric permutations is always upper bounded by the complexity of $\mathcal{T}(P)$. The first term is interesting because it depends only on $k$ (and improves Wenger's bound). Still, the only known lower bound on this quantity is $\Omega(k^2)$, even for the restricted case of lines in $\mathcal{L}$, and it would be very interesting to show that this is also an upper bound in the special case of collections of pairwise disjoint convex objects in $\mathbb{R}^3$, one of which is a line. See Section H for details.

Due to lack of space, many details, such as missing proofs of lemmas and theorems, are delegated to the appendices.

## 2 Preliminaries

Let $\mathcal{P} = \{P_1, \ldots, P_k\}$ be a collection of $k$ convex polyhedra in $\mathbb{R}^3$ with a total of $n$ facets, and let $\ell_0$ be a fixed line. We further assume that the polyhedra of $\mathcal{P}$ and $\ell_0$ are in general position.[4] Without loss of generality, we take $\ell_0$, for the time being, to be the $z$-axis.

Let $\mathcal{L} = \mathcal{L}_{\ell_0}$ denote the space of all lines that pass through $\ell_0$ (other than $\ell_0$ itself). Lines in $\mathcal{L}$ have three degrees of freedom, and we represent each (directed) line $\ell \in \mathcal{L}$ by a triple $(\theta(\ell), \varphi(\ell), z(\ell))$, where $z(\ell)$ is the $z$-coordinate of the intercept of $\ell$ at $\ell_0$, and $(\theta(\ell), \varphi(\ell))$ are the spherical coordinates of the orientation of $\ell$. Clearly, all lines $\ell$ with $\theta(\ell) = \theta$ lie in the plane through $\ell_0$ at $xy$-orientation $\theta$; we denote this plane by $\Pi_\theta$. See Figure 1 (left). The intersection of $\Pi_\theta$ with a polyhedron $P$ is the polygon $P(\theta) = P \cap \Pi_\theta$ (if it is not empty).

---

[3]The reason for the index 4 is that $\beta_4(k) = \Theta(\frac{\lambda_4(k)}{k})$, where $\lambda_4(k)$ is the maximum length of Davenport-Schinzel sequences of order 4 on $k$ symbols; see [27] and below.

[4]In particular, we assume that at most three edges, or a vertex and an edge of the given polyhedra, admit a common transversal through $\ell_0$, and that no edge of any polyhedron is coplanar with $\ell_0$ or with a facet of another polyhedron.

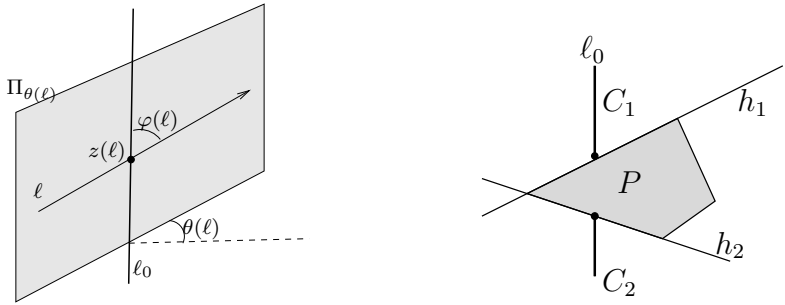

Figure 1: (Left) Representing an oriented line $\ell \in L$. The plane $\Pi_{\theta(\ell)}$ contains $\ell$, $(\theta(\ell), \varphi(\ell))$ are the spherical coordinates of the orientation vector of $\ell$, and $z(\ell) = \ell \cap \ell_0$ is the $z$-intercept of $\ell$. (Right) We separate the components $C_1$ and $C_2$ of $\ell_0 \setminus P$ by the planes $h_1$ and $h_2$, respectively, each containing a facet of $P$.

We define, for each polyhedron $P \in \mathcal{P}$, a pair of (partial) bivariate functions $\sigma_P^+$ and $\sigma_P^-$, over the $\theta\varphi$-domain, so that $\sigma_P^-(\theta, \varphi)$ (resp., $\sigma_P^+(\theta, \varphi)$) is the $z$-intercept (at $\ell_0$) of the line whose orientation has spherical coordinates $(\theta, \varphi)$ and which is tangent to $P$ from below (resp., above).

For each $P \in \mathcal{P}$, the graphs of $\sigma_P^+$ and $\sigma_P^-$, are $(\theta, \varphi)$-monotone surfaces, representing tangents to the upper and lower portions of $\partial P$, respectively. With an appropriate re-parametrization (e.g., replacing $\theta$ with $\tan \frac{\theta}{2}$, and $\varphi$ with $\cot \varphi$), each surface $\sigma_P^-$ (resp., $\sigma_P^+$) consists of monotone semi-algebraic surface patches, each of which is a graph of a partially-defined function, representing lower (resp., upper) tangents to $P$ at a fixed edge of its lower (resp., upper) boundary.

The set $\mathcal{T}_{\ell_0}(\mathcal{P})$ of all transversals to $\mathcal{P}$ (in $\mathcal{L}$) is then the *sandwich region*

$$\left\{ (\theta, \varphi, z) \in \mathcal{L} \mid \max_{P \in \mathcal{P}} \sigma_P^-(\theta, \varphi) \leq z \leq \min_{P \in \mathcal{P}} \sigma_P^+(\theta, \varphi) \right\} \tag{1}$$

between the upper envelope $E_U = \max_{P \in \mathcal{P}} \sigma_P^-$ of the functions $\sigma_P^-$ and the lower envelope $E_L = \min_{P \in \mathcal{P}} \sigma_P^+$ of the functions $\sigma_P^+$.

Following [7, 24] (see also the introduction), we define an *extremal line* to be a line $\ell$ tangent to some polyhedra of $\mathcal{P}$ at a set $A$ of respective vertices and edges, so that $\ell$ cannot be continuously moved (within $\mathcal{L}$) while remaining transversal to the elements of $A$. An *extremal stabbing line* is an extremal line, which is also a transversal of $\mathcal{P}$. Every vertex of $\mathcal{T}_{\ell_0}(\mathcal{P})$ corresponds to an extremal stabbing line, and vice versa; see, e.g., [24]. The following theorem was proven by Brönnimann et al. [7].

**Theorem 2.1.** *Let $P$ and $Q$ be two convex polyhedra (in general position with respect to each other, and to $\ell_0$) in $\mathbb{R}^3$, having $n_P$ and $n_Q$ facets, respectively. Then the arrangement of the tangency surfaces $\sigma_P^+$, $\sigma_P^-$, $\sigma_Q^+$, and $\sigma_Q^-$ has combinatorial complexity $O(n_P + n_Q)$, and it can be computed in $O((n_P + n_Q) \log(n_P + n_Q))$ time. This implies that there are $O(n_P + n_Q)$ pairs of edges, one of $P$ and one of $Q$, which admit common tangent lines to $P, Q$ through them which belong to $\mathcal{L}$. In particular, the combinatorial complexity of $\mathcal{T}_{\ell_0}(\{P, Q\})$ is $O(n_P + n_Q)$, and it can be computed in $O((n_P + n_Q) \log(n_P + n_Q))$ time.*

We say that a vertex or an edge $\xi$ of $\mathcal{T}_{\ell_0}(\mathcal{P})$ is *defined* by a set of polyhedra $\mathcal{P}' \subset \mathcal{P}$ if $\mathcal{P}'$ is a minimal set of polyhedra such that $\xi$ is present in $\mathcal{T}_{\ell_0}(\mathcal{P}')$. Assuming general position of $\mathcal{P} \cup \{\ell_0\}$, each vertex $v$ of $\mathcal{T}_{\ell_0}(\mathcal{P})$ is defined by a unique set of between one and three polyhedra.

By Theorem 2.1, any two polyhedra $P, Q \in \mathcal{P}$, having, respectively, $n_P$ and $n_Q$ facets, define $O(n_P + n_Q)$ features of $\mathcal{T}_{\ell_0}(\mathcal{P})$, each of which is the locus of lines tangent to $P$ and $Q$ at two specific boundary edges. Summing over all pairs of polyhedra in $\mathcal{P}$, we obtain a bound of $O(nk)$ on the number of features of this kind defined by at most two polyhedra. Any other feature can be charged to a vertex of $\mathcal{T}_{\ell_0}(\mathcal{P})$, so that no vertex is charged more than $O(1)$ times. Therefore, it is sufficient to bound the number of vertices of $\mathcal{T}_{\ell_0}(\mathcal{P})$ which are defined by three polyhedra; each such vertex is an extremal stabbing line (in $\mathcal{L}$) which is a common tangent to three polyhedra of $\mathcal{P}$, at (the relative interiors of) three edges, one of each polyhedron. The additional complexity $O(nk)$ will be subsumed in the bound that we will obtain.

Let $P$ and $Q$ be a pair of polyhedra of $\mathcal{P}$, and let $\zeta$ be a boundary edge of $\mathcal{T}_{\ell_0}(\{P, Q\})$, contained in the common intersection of two semi-algebraic patches $\sigma_e, \sigma_{e'}$, where $\sigma_e$ (resp., $\sigma_{e'}$) is contained in $\sigma_P^+ \cup \sigma_P^-$ (resp., $\sigma_Q^+ \cup \sigma_Q^-$) and represents oriented lines (in $\mathcal{L}$) tangent to $P$ (resp., $Q$) at $e$ (resp., $e'$). That is, $\zeta$ represents (i.e., is the trace of) a maximal connected set of lines, that are tangent to $P$ at $e$, and to $Q$ at $e'$. Note that $\zeta$ is a connected portion of the locus of lines that pass through three fixed lines, namely, $\ell_0$, and the two lines supporting $e$ and $e'$, respectively. In general position, this locus is commonly referred as a *regulus*, whose lines have only one degree of freedom, and trace (a portion of) a ruled surface in $\mathbb{R}^3$, which is either a hyperbolic paraboloid or a 1-sheeted hyperboloid; see [9, 29] for more details on reguli.

In the restricted context of this paper, a *regulus* denotes a maximal connected set of oriented lines in $\mathcal{L}$, that are tangent to two fixed polyhedra of $\mathcal{P}$, at two fixed edges, one of each polyhedron. In particular, each regulus represents some boundary edge of $\mathcal{T}_{\ell_0}(\{P, Q\})$, for some pair of distinct polyhedra $P, Q \in \mathcal{P}$. Lemma 2.1, together with the follow-up discussion, imply that the polyhedra of $\mathcal{P}$ define a total of $O(nk)$ such reguli.

**Transversals parallel to a fixed plane.** Let $h$ be a fixed plane in $\mathbb{R}^3$. Denote by $\mathcal{L}_h$ the space of lines passing through $\ell_0$ and parallel to $h$. Clearly, lines in $\mathcal{L}_h$ have only two degrees of freedom, and, if $h$ is generic, any extremal stabbing line within $\mathcal{L}_h$ is tangent to at most two polyhedra at a corresponding pair of edges. We establish the following lemma, which we need in our analysis.

**Lemma 2.2.** *Let $h$ be a fixed plane in $\mathbb{R}^3$. Then the number of extremal stabbing lines of $\mathcal{P}$ within the space $\mathcal{L}_h$, as defined above, is $O(n\beta_4(k)) = O(n \cdot 2^{\alpha(k)})$.*

**Separating convex bodies in $\mathbb{R}^3$.** We denote by $SS^2$ the unit sphere in 3-space centered at the origin. For each oriented line $\ell$ in 3-space, we denote its orientation by $\vec{d}(\ell)$, and represent it as a point on $SS^2$, with spherical coordinates $(\theta(\ell), \varphi(\ell))$. For a plane $h \subset \mathbb{R}^3$, we denote by $C_h \subset SS^2$ the great circle obtained by intersecting $SS^2$ with the plane parallel to $h$ and containing the origin; equivalently, $C_h$ is the locus of all orientations on $SS^2$ that are parallel to $h$. The following (easy) lemma has been proven by Wenger [30].

**Lemma 2.3.** *Let $P$ and $Q$ be a pair of disjoint convex bodies in $\mathbb{R}^3$, and let $h$ be a plane which separates them. Then $C_h$ partitions $SS^2$ into a pair of hemispheres $SS_h^+$ and $SS_h^-$, such that, for any (oriented) common transversal $\ell$ of $P$ and $Q$, $\ell$ stabs $P$ before (resp., after) $Q$ if and only if $\vec{d}(\ell) \in SS_h^+$ (resp., $\vec{d}(\ell) \in SS_h^-$).*

## 3 The Complexity of $\mathcal{T}(\mathcal{P})$ and Its Construction

We first establish Theorem 3.1 that bounds the complexity of $\mathcal{T}_{\ell_0}(\mathcal{P})$. In order to get the bound on the complexity of $\mathcal{T}_{\ell_0}(\mathcal{P})$ to depend on $k$, we reduce the global problem, involving a sandwich region in the 3-dimensional space $\mathcal{L}$, to a collection of 2-dimensional problems. The difficulty is that a naive reduction of this sort yields subproblems in which the relevant portion of $\mathcal{T}_{\ell_0}(\mathcal{P})$ is *not* a sandwich region. Our solution uses a more involved approach to ensure that the resulting subproblems do have a sandwich structure, but this requires more careful and somewhat intricate analysis. Then we use Theorem 3.1 to establish a bound on the complexity of $\mathcal{T}(\mathcal{P})$.

**Theorem 3.1.** *Let $\mathcal{P}$ be a set of $k$ convex polyhedra in $\mathbb{R}^3$ with a total of $n$ facets, and let $\ell_0$ be a fixed line (all in general position). Then the number of extremal stabbing lines to $\mathcal{P}$ in the set $\mathcal{L}$ of lines passing through $\ell_0$ is $O(nk^{1+\varepsilon})$, for any $\varepsilon > 0$. One can construct collections of $k$ convex polyhedra in general position (together with $\ell_0$), with a total of $n$ facets, for arbitrarily large values of $k$ and $n$, for which the complexity of the stabbing region is $\Omega(nk)$.*

*Proof.* For each polyhedron $P \in \mathcal{P}$, we separate each connected component of $\ell_0 \setminus P$ from $P$ by a plane (there are at most two such components, and therefore at most two corresponding planes; if $\ell_0 \cap P = \emptyset$, there is a single separating plane which is parallel to $\ell_0$). See Figure 1 (right) for an illustration. Let $H$ be the resulting set of at most $2k$ separating planes. These planes intersect $\ell_0$ in at most $2k$ points, partitioning it into a collection $\mathcal{I}$ of up to $2k+1$ open "atomic" intervals, so that, for each interval $I \in \mathcal{I}$ and for each polyhedron $P \in \mathcal{P}$, either $I$ is fully contained in $P$ or $I$ is disjoint from $P$.

Let $C_H$ denote the collection of the great circles $C_h \subset SS^2$, for $h \in H$, as defined in the previous section, and let $\mathcal{A}(C_H)$ denote the arrangement that they form on the sphere $SS^2$. The construction and Lemma 2.3 imply that for each cell $D$ of $\mathcal{A}(C_H)$, and for each $I \in \mathcal{I}$, there exists a partition of $\mathcal{P}$ into three subsets $\mathcal{P}_0 = \mathcal{P}_0(D, I)$, $\mathcal{P}^- = \mathcal{P}^-(D, I)$, and $\mathcal{P}^+ = \mathcal{P}^+(D, I)$ with the following properties. For any oriented stabbing line $\ell \in \mathcal{L}$ which emanates from a point on $I$ and has orientation in $D$, the point $q = \ell \cap \ell_0$ lies in $P$ if $P \in \mathcal{P}_0$, before $\ell \cap P$ along (the oriented) $\ell$ if $P \in \mathcal{P}^-$, and after $\ell \cap P$ along (the oriented) $\ell$ if $P \in \mathcal{P}^+$.

We now fix an edge $e_0$ on the boundary of some polyhedron $P_0 \in \mathcal{P}$, and denote by $\mathcal{L}[e_0]$ the set of *oriented* lines which pass through $\ell_0$ and are tangent to $P_0$ at (the interior of) $e_0$. (Assuming general position of $\mathcal{P}$ and $\ell_0$, we can exclude extremal stabbing lines that pass through one of the endpoints of $e_0$, or overlap a facet of $P_0$. As argued in Section 2 the number of such "degenerate" stabbers is only $O(nk)$. We parametrize these lines by the appropriately rotated spherical coordinate system, in which the line $\ell_{e_0}$ supporting $e_0$ is the $z$-axis (with a fixed assigned orientation), and the $xz$-plane supports one of the facets of $P_0$ incident to $e_0$. It is easy to check that, in general position, this is a unique parametrization. Thus, in the new system, which, for convenience, we continue to denote by $(\theta, \varphi)$, the angle $\theta$ encodes the orientation of a plane $\tilde{\Pi}_\theta$ containing $e_0$ and rotating about it. However, we are only interested in values of $\theta$ for which $\tilde{\Pi}_\theta$ is tangent to $P_0$ at $e_0$. Thus, if $\theta_0$ denotes $\pi$ minus the dihedral angle of $P_0$ at $e_0$, then we can restrict $\theta$ to lie in the union of the two antipodal angular ranges $(0, \theta_0)$, and $(\pi, \theta_0 + \pi)$. For simplicity, we only consider lines in $\mathcal{L}[e_0]$ whose orientation $(\theta, \varphi)$ satisfies $0 < \theta < \theta_0$. See Figure 2 for an illustration.

For a fixed $0 < \theta < \theta_0$, we intersect each $P \in \mathcal{P}$ with the plane $\tilde{\Pi}_\theta$, thus obtaining a set $\tilde{\mathcal{P}}(\theta)$ of (possibly empty) convex polygons of the form $\tilde{P}(\theta) := P \cap \tilde{\Pi}_\theta$. In particular, we have $\tilde{P}_0(\theta) \equiv e_0$, for all $\theta \in (0, \theta_0)$.

Let $\ell_0(\theta)$ denote the point $\ell_0 \cap \tilde{\Pi}_\theta$. By the general position assumption, $e_0$ is not co-planar with $\ell_0$, so the point $\ell_0(\theta)$ is well-defined, except for the unique orientation $\theta^*_{e_0}$, if it exists, at which $\tilde{\Pi}_\theta$ is parallel to $\ell_0$, which we ignore here.

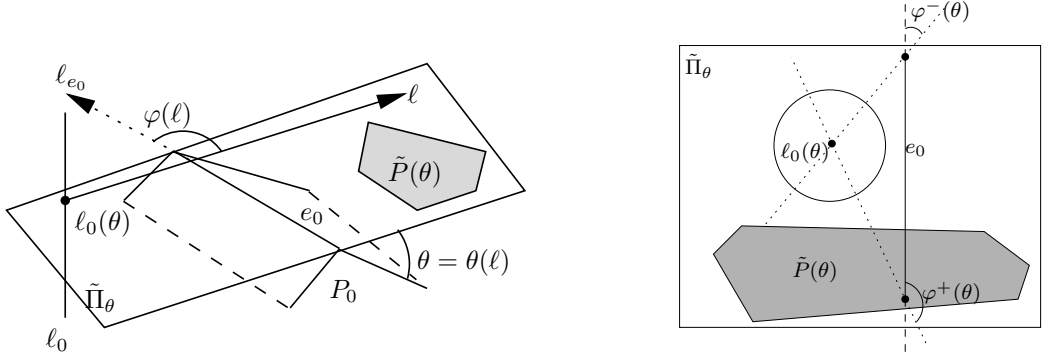

Figure 2: Representing lines in $\mathcal{L}[e_0]$: A side view (left), and the cross-section within the plane $\tilde{\Pi}_\theta$ (right).

Now with all these preparations, the portion under consideration of $\mathcal{L}[e_0]$, which, for simplicity of presentation, we continue to denote by $\mathcal{L}[e_0]$, is represented as the region[5]

$$W = \left\{ (\theta, \varphi) \mid 0 < \theta < \theta_0, \varphi^-(\theta) < \varphi < \varphi^+(\theta) \right\},$$

where $\varphi^-(\theta), \varphi^+(\theta)$ are the $\varphi$-coordinates of the respective lines passing through $\ell_0(\theta)$ and through each of the endpoints of $e_0$. See Figure 2 (right) for an illustration.

Next we fix a cell $D$ of $\mathcal{A}(C_H)$ (defined at the beginning of the proof), and consider the subset $\mathcal{L}[e_0, D]$ of those lines $\ell \in W$ with orientation in $D$. (If $\mathcal{L}[e_0, D]$ is empty then we ignore $D$.)

For each polyhedron $P \in \mathcal{P}$, we define two functions $\gamma_P^-(\theta), \gamma_P^+(\theta)$, for $\theta \in (0, \theta_0)$, as follows. Let $I \in \mathcal{I}$ be the segment which contains $\ell_0(\theta)$, then we define $\gamma_P^-(\theta)$ ($\gamma_P^+(\theta)$) to be the minimum (resp., maximum) value of $\varphi \in (\varphi^-(\theta), \varphi^+(\theta))$ such that (a) the line $\ell$ with representation $(\theta, \varphi)$ intersects $P$, and (b) the order of $\ell_0 \cap \ell = \ell_0(\theta)$ and $P \cap \ell$ along $\ell$ is as prescribed by $D$ and $I$;[6] see Figure 3 (right). A more detailed definition of $\gamma_P^+, \gamma_P^-$ is given in Appendix B. Although the definition is fairly straightforward, it provides the central tool for expressing the stabbing region in $\mathcal{L}[e_0, D]$ as a *sandwich region*; see Lemma 3.3.

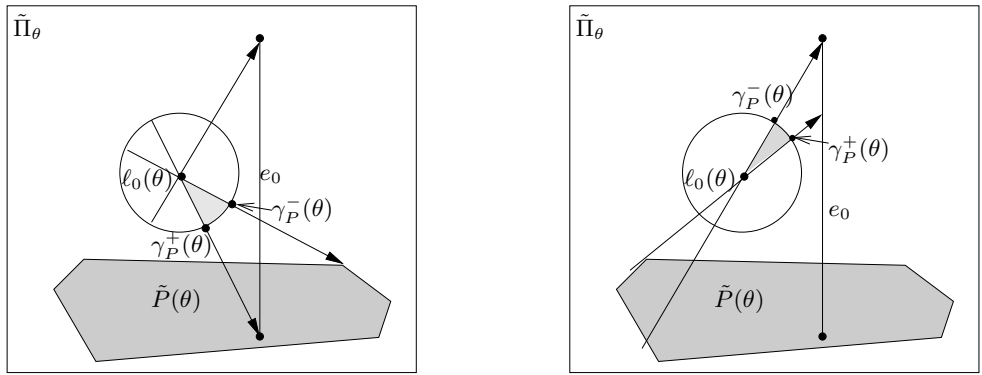

Figure 3: The interval $(\gamma_P^-(\theta), \gamma_P^+(\theta))$ when $P \in \mathcal{P}^+$ (left), and $P \in \mathcal{P}^-$ (right).

---
[5]The inequalities are sharp since, lines in $\mathcal{L}[e_0]$ can neither overlap a facet of $P_0$ incident to $e_0$ nor pass through any of the endpoints of $e_0$; see the definition of $\mathcal{L}[e_0]$.

[6]Namely, $\ell_0(\theta)$ is contained in $P$ (and thus $\tilde{P}(\theta)$), if $P \in \mathcal{P}_0(D, I)$; and $P$ (and, therefore, $\tilde{P}(\theta) \cap \ell$) lies before (resp., after) $\ell_0(\theta)$ along $\ell$, if $P \in \mathcal{P}^-(D, I)$ (resp., $P \in \mathcal{P}^-(D, I)$).

It follows that, for $\theta \in (0, \theta_0)$, the line $\ell$ having representation $(\theta, \gamma_P^-(\theta))$ $((\theta, \gamma_P^+(\theta)))$ is one of: (a) a tangent to $\tilde{P}(\theta)$ passing through $\ell_0(\theta)$ in the plane $\tilde{\Pi}_\theta$, or (b) a line which connects $\ell_0(\theta)$ with one of the endpoints of $e_0 = \tilde{P}_0(\theta)$. With an appropriate, standard re-parametrization of $\theta$, $\varphi$, as above, the graph of each of the functions $\gamma_P^-$, $\gamma_P^+$ is piecewise algebraic, composed of algebraic arcs of constant maximum degree. Each arc either represents lines in $\mathcal{L}[e_0]$ that are tangent to $P$ at some fixed edge (each such arc is a trace, on $SS^2$, of a *regulus* of lines, defined in Section 2), or it represents lines tangent to $P_0$ at one of the endpoints of $e_0$. Since by assumption all lines in $\mathcal{L}[e_0]$ have $\varphi \in (\varphi^-(\theta), \varphi^+(\theta))$, the arcs of the latter type are redundant and, therefore, can be safely ignored in the subsequent analysis. It follows that any pair of the remaining arcs, corresponding to a pair of edges $e, e'$ of distinct respective polyhedra $P, P'$, intersect at most twice, since these intersections represent lines that pass through four fixed lines: $\ell_0$, and the lines containing the edges $e_0$, $e$, and $e'$; see [29].

Formally, in Appendix B we prove the following lemma.

**Lemma 3.2.** *Let $P$ be a polyhedron of $\mathcal{P} \setminus \{P_0\}$. Then each algebraic subarc $\zeta$ of $\gamma_P^+$, defined by some fixed edge $e$ of $P$, is (a trace of) a regulus defined by $e_0$ and $e$. Namely, $\zeta$ represents a maximal connected set of (oriented) lines in $\mathcal{L}$, that are tangent to $P_0$ at $e_0$, and to $P$ at $e$.*

Now the discussion following Theorem 2.1 implies that the overall number of distinct algebraic arcs that comprise the graphs of the functions $\gamma_P^+$, $\gamma_P^-$, over all possible edges $e_0$, polyhedra $P$, and cells $D$, is $O(nk)$ (where a single arc may traverse several cells $D \in \mathcal{D}$). Indeed, for a given pair of distinct polyhedra $P_1, P_2 \in \mathcal{P}$, the regulus of lines tangent to $P_1$, at a fixed edge $e_1$, and to $P_2$, at a fixed edge $e_2$, can show up (with different parametrizations) only in the subspaces $\mathcal{L}[e_1], \mathcal{L}[e_2]$ of $\mathcal{L}$, although it can appear in the subsets $\mathcal{L}[e_1, D]$ of $\mathcal{L}[e_1]$ for several cells $D$, and similarly for $e_2$.

**Lemma 3.3.** *Let $e_0$, $D$, and $P$ be as above, and let $\ell$ be an oriented line in $\mathcal{L}[e_0, D]$, with (the local) spherical coordinates $(\theta, \varphi)$. Then $\ell$ stabs $P$ if and only if*

$$\gamma_P^-(\theta) \leq \varphi \leq \gamma_P^+(\theta). \tag{2}$$

It follows from Lemma 3.3 that for a fixed $D$ and $e_0$, a line $\ell \in \mathcal{L}[e_0, D]$, with coordinates $(\theta, \varphi)$, is a transversal to $\mathcal{P}$ if and only if

$$\max_{P \in \mathcal{P}} \gamma_P^-(\theta) \leq \varphi \leq \min_{P \in \mathcal{P}} \gamma_P^+(\theta). \tag{3}$$

**Corollary 3.4.** *Let $e_0$ and $D$ be as above, and let $\ell$ be an oriented extremal stabbing line to $\mathcal{P}$ in $\mathcal{L}[e_0, D]$, with coordinates $(\theta, \varphi)$. Then the point $(\theta, \varphi)$ corresponds to a vertex on the boundary of the sandwich region given by (3).*

Hence, the number of extremal stabbing lines in $\mathcal{L}[e_0, D]$ is upper bounded by the complexity of the sandwich region given by (3). Since this region is formed by the graphs of $O(k)$ functions, consisting of some number, $\Gamma[e_0, D]$, of algebraic arcs, each pair of which intersect at most twice, it follows (as in the proof of Lemma 2.2) from [27, Theorem 1.4] that the complexity of the sandwich region is $O(\Gamma[e_0, D]\beta_4(k))$.

In other words, if we fix $D$, and sum this bound over all edges $e_0$ of the polyhedra of $\mathcal{P}$, we get an overall bound of $O(nk\beta_4(k))$. Unfortunately, multiplying this bound by the number, $O(k^2)$, of cells $D$, yields an upper bound of $O(nk^3\beta_4(k))$, which is much too large. (Here we face the difficulty that arcs may appear in several cells $D$.)

In order to keep the overall bound close to $O(nk)$, we note that we can replace $\Gamma[e_0, D]$ by the number $R[e_0, D]$ of those arcs that actually show up on the boundary of the sandwich region (3), at points which represent lines in $\mathcal{L}[e_0, D]$. That is, we exclude arcs which either do not appear on the boundary of the sandwich region at all, or appear there but only at (lines represented by) orientations outside $D$. Our goal is to bound $\sum_{e_0, D} R[e_0, D]$. See Figure 4 for an illustration.

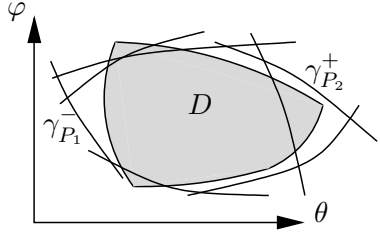

Figure 4: The sandwich region in $\mathcal{L}[e_0, D]$, as given by (3), and $D$, drawn on the unit sphere $SS^2$. Note that $\gamma_{P_1}^-$ and $\gamma_{P_2}^+$ appear on the boundary of the sandwich region, but not within $D$, so their arcs are not counted in $\mathcal{R}[e_0, D]$.

To recap, the preceding analysis implies that the number $N[e_0, D]$ of extremal stabbing lines in $\mathcal{L}[e_0, D]$ satisfies
$$N[e_0, D] = O\left(R[e_0, D]\beta_4(k)\right). \tag{4}$$
Put $N = \sum_{e_0, D} N[e_0, D]$ and $R = \sum_{e_0, D} R[e_0, D]$, where the sums extend over all choices of edges $e_0$ of the polyhedra in $\mathcal{P}$ and all cells $D$ of $\mathcal{A}(C_H)$. Denote by $N(n, k)$ (resp., $R(n, k)$) the maximum value of $N$ (resp., $R$), over all possible choices of a collection $\mathcal{P}$ of $k$ convex polyhedra with a total of $n$ facets, and of a fixed line $\ell_0$.

By (4), we have $N(n, k) = O(R(n, k)\beta_4(k))$. Now the last step of the analysis is deriving a recurrence formula for $N(n, k)$ which we obtain by proving that
$$R(n, k) = O\left(nk + tnk\beta_4(k) + \frac{1}{t} \cdot t^3 N\left(\frac{n}{t}, \frac{k}{t}\right)\right). \tag{5}$$

for any integer threshold $t > 0$. The desired bound is the solution of this recurrence. We derive (5) using the slide-and-charge scheme described, e.g., in [26]; see also [27]. Specifically, we say that the level of a line $\ell \in \mathcal{L}$ is $j$ if $\ell$ stabs all but $j$ polyhedra of $\mathcal{P}$. Now, fix a polyhedron $P_0$ and an edge $e_0$ of $\partial P_0$. Let $\zeta$ be an arc which is counted by $R[e_0, D]$, for at least one cell $D \in \mathcal{D}$. As argued above, there is a total of $O(nk)$ such arcs, over all choices of $P_0$ and $e_0$. If $\zeta$ is counted $R[e_0, D]$, for more than one cell $D \in \mathcal{D}$, we charge each of its appearances (except for the first one) either to (i) an extremal line in $\mathcal{L}_h$ (see Section 2), at level $\leq t$, where $h$ is one of the planes such that the great circle $C_h$ appears on the boundary of $D$, or to (ii) $t$ extremal stabbing lines in $\mathcal{L}[e_0, D]$, at level $\leq t$. Combining Lemma 2.2 with the general randomized technique of Clarkson and Shor [10], we get that the number of extremal stabbing lines at level $\leq t$ in $\mathcal{L}_h$ is bounded by $O(nt\beta_4(k))$, and the number of extremal stabbing lines at level $\leq t$ in all $O(k)$ planes $h \in H$ is $O(nkt\beta_4(k))$. The details are in appendix C. The lower bound is proved in appendix D. $\square$

**Theorem 3.5 (The general $\mathcal{T}(\mathcal{P})$).** *Let $\mathcal{P}$ be a collection of $k$ convex polyhedra in $\mathbb{R}^3$ with a total of $n$ facets. Then the set $\mathcal{T}(\mathcal{P})$ of line transversals of $\mathcal{P}$ has complexity $O(n^2 k^{1+\varepsilon})$, for any $\varepsilon > 0$.*

**Theorem 3.6 (Algorithm $\mathcal{T}_{\ell_0}(\mathcal{P})$).** *Let $\mathcal{P}$ be a set of $k$ convex polyhedra with a total of $n$ facets, and let $\ell_0$ be a fixed line (all in general position). Then one can compute the stabbing region $\mathcal{T}_{\ell_0}(\mathcal{P})$ of $\mathcal{P}$, within the set $\mathcal{L}$ of lines passing through $\ell_0$, in $O(nk^{1+\varepsilon}\log n)$ randomized expected time, for any $\varepsilon > 0$.*

The following theorem follows from Theorem 3.6 and Theorem 3.5.

**Theorem 3.7 (Algorithm $\mathcal{T}(\mathcal{P})$).** *Let $\mathcal{P}$ be a set of $k$ convex polyhedra with a total of $n$ facets. Then one can compute (the boundary representation of) the stabbing region $\mathcal{T}(\mathcal{P})$ in $O(n^2 k^{1+\varepsilon}\log n)$ randomized expected time, for any $\varepsilon > 0$.*

## 4 Extensions

In Appendix G we prove slightly improved upper bounds on the complexity of the stabbing region $\mathcal{T}_{\ell_0}(\mathcal{P})$, in the following two cases: the case where all polyhedra of $\mathcal{P}$ are disjoint from $\ell_0$, and the case where the polyhedra of $\mathcal{P}$ are all unbounded in one of the two directions parallel to $\ell_0$. In both cases, we also provide deterministic algorithms for efficiently computing $\mathcal{T}_{\ell_0}(\mathcal{P})$. The first result is established in Theorem 4.1 and the second is a corollary of Theorem 4.2.

**Theorem 4.1.** *Let $\mathcal{P}$ be a set of $k$ pairwise disjoint convex polyhedra with a total of $n$ facets, and let $\ell_0$ be a fixed line disjoint from all polyhedra of $\mathcal{P}$ (all in general position). Then the complexity of the stabbing region $\mathcal{T}_{\ell_0}(\mathcal{P})$ of $\mathcal{P}$ in the space $\mathcal{L}$ of lines passing through $\ell_0$ is $O((nk+k^3)\beta_4(k))$. Moreover, one can compute $\mathcal{T}_{\ell_0}(\mathcal{P})$ in $O((nk+k^3)(\log n + \alpha(k)\log k))$ deterministic time.*

**Theorem 4.2.** *Let $\mathcal{P}$ be a collection of $k$ pairwise disjoint convex polyhedra in $\mathbb{R}^3$, having a total of $n$ facets, and let $\ell_0$ be a fixed line. The number of vertices of the upper envelope $E_U$ of the partially-defined functions $\sigma_P^-$, for $P \in \mathcal{P}$ (as defined in Section 2), is $O(nk\beta_4(k))$. Moreover, there is a deterministic algorithm which computes $E_U$ in $O(nk(\log n + \alpha(k)\log k))$ time.*

Using our techniques we were also able to probe the following result about geometric permutations.

**Theorem 4.3.** *Let $\mathcal{P}$ be a collection of $k$ pairwise disjoint convex objects in $\mathbb{R}^3$, one of which is a line $\ell_0$. Then the number of geometric permutations induced by $\mathcal{P}$ is $O(k^3)$.*

## 5 Conclusion

In this paper we obtained an improved bound on the combinatorial complexity of the set $\mathcal{T}(\mathcal{P})$ of line transversals to a collection $\mathcal{P}$ of $k$ convex polyhedra in $\mathbb{R}^3$ with a total of $n$ facets, and showed how to compute $\mathcal{T}(\mathcal{P})$ in comparable randomized expected time. We reduce this general problem to the restricted instance in which line transversals of $\mathcal{P}$ are constrained to pass through some fixed line $\ell_0$. Specializing to this restricted instance, we obtain *nearly tight* bounds on the complexity of the stabbing region $\mathcal{T}_{\ell_0}(\mathcal{P})$ within the space $\mathcal{L}$ of lines that pass through $\ell_0$. Our analysis successfully combines the slide-and-charge scheme from [26] with the method of separating planes, that was previously used in [30] to study geometric permutations; also see [5].

There are several challenging open problems for further research, including the problem of closing the gap between the lower and the upper bounds for the complexity of the stabbing region $\mathcal{T}(\mathcal{P})$, as given in Section 3.

## A Proof of Lemma 2.2

**Lemma A.1.** *Let $h$ be a fixed plane in $\mathbb{R}^3$. Then the number of extremal stabbing lines of $\mathcal{P}$ within the space $\mathcal{L}_h$, as defined above, is $O(n\beta_4(k)) = O(n \cdot 2^{\alpha(k)})$.*

*Proof.* Assume first that $h$ is not parallel to $\ell_0$. Then lines in $\mathcal{L}_h$ can be parametrized by their two coordinates $(\theta, z)$, where the third coordinate $\varphi = \varphi_h(\theta)$ depends only on $\theta$; with an appropriate re-parametrization, e.g., replacing $\theta$ by $\tan\frac{\theta}{2}$, and $\varphi$ by $\cot\varphi$, the function $\varphi_h(\theta)$ is a low-degree algebraic function. Hence, using (1), the stabbing region of $\mathcal{P}$ within $\mathcal{L}_h$ is the set

$$\left\{ (\theta, z) \mid \max_{P \in \mathcal{P}} \sigma_P^-(\theta, \varphi_h(\theta)) \leq z \leq \min_{P \in \mathcal{P}} \sigma_P^+(\theta, \varphi_h(\theta)) \right\}.$$

That is, the transversal space within $\mathcal{L}_h$ is the sandwich region between the lower envelope of the $k$ *univariate* functions $\sigma_P^+(\theta, \varphi_h(\theta))$, and the upper envelope of the $k$ functions $\sigma_P^-(\theta, \varphi_h(\theta))$, for $P \in \mathcal{P}$. The graphs of these functions are piecewise algebraic, where each piece represents lines tangent to some specific polyhedron at some specific edge of its boundary. The overall number of these pieces (subarcs) is thus $O(n)$, and any pair of them intersect at most twice. Indeed, any such intersection represents a line $\ell$ in $\mathcal{L}$, parallel to $h$, which intersects the two polyhedra edges $e_1$ and $e_2$ corresponding to those subarcs. Hence, $\ell$ is a line that passes through $\ell_0$, the two lines containing $e_1$ and $e_2$, and the line at infinity in $h$. Assuming general position, there are at most two such lines $\ell$ [29].

Consider next the case where $h$ is parallel to $\ell_0$. We can assume that $h$ contains $\ell_0$, and thus equals to a plane $\Pi_{\theta_h}$, for some fixed $0 \leq \theta_h \leq 2\pi$. That is, the transversal space within $\mathcal{L}_h$ is the sandwich region between the lower envelope of the $k$ univariate functions (of $\varphi$) $\sigma_P^+(\theta_h, \varphi)$, and the upper envelope of the $k$ functions $\sigma_P^-(\theta_h, \varphi)$, for $P \in \mathcal{P}$. Again, with an appropriate re-parametrization (e.g., replacing $\varphi$ by $\cot\varphi$), the graphs of those functions are piecewise algebraic, where each piece represents lines tangent to some specific polygon $P(\theta_h)$ at some specific vertex of its boundary; in this case any pair of these pieces intersect at most once—the intersection corresponds to the unique line in $\Pi_{\theta_h}$ that passes through the two respective vertices. See [17] for a similar analysis. The overall number of these pieces is, as above, $O(n)$.

In both cases, the lemma follows from the upper bound on the complexity of the sandwich region defined by $O(k)$ piecewise-algebraic curves in $\mathbb{R}^2$, which are composed of a total of $O(n)$ algebraic arcs, each pair of which intersect at most twice. Recalling our definition of $\beta_4(k)$, this bound is $O(\frac{n}{k}\lambda_4(k)) = O(n\beta_4(k))$, where $\lambda_4(k) = \Theta(k \cdot 2^{\alpha(k)})$ is the maximum length of Davenport-Schinzel sequences of order 4 on $n$ symbols; see [27, Theorem 1.4]. □

## B  Details omitted from the proof of Theorem 3.1

**Formal definition of $\gamma_P^-(\theta)$ and $\gamma_P^+(\theta)$.** For each polyhedron $P \in \mathcal{P}$, we define two functions $\gamma_P^-(\theta)$, $\gamma_P^+(\theta)$, for $\theta \in (0, \theta_0)$, as follows.

Fixing the value of $\theta$, we get a fixed plane $\tilde{\Pi}_\theta$, and a fixed intercept $\ell_0(\theta) = \ell_0 \cap \tilde{\Pi}_\theta$. Let $I$ denote the atomic interval of the partition $\mathcal{I}$ containing $\ell_0(\theta)$. Let $\mathcal{P}_0$, $\mathcal{P}^-$, $\mathcal{P}^+$ denote the corresponding partition of $\mathcal{P}$ induced by $D$ and $I$.

If $P \in \mathcal{P}_0$ then any line $\ell \in \mathcal{L}[e_0, D]$ with $\theta(\ell) = \theta$ stabs $P$, and none of these lines can be tangent to $P$. So we set in this case $\gamma_P^-(\theta) = \varphi^-(\theta)$ and $\gamma_P^+(\theta) = \varphi^+(\theta)$. Technically, this prevents $P$ from affecting the region of transversals in $\mathcal{L}[e_0, D]$.

Suppose then that $P \in \mathcal{P}^+$. The set of $\varphi \in (\varphi^-(\theta), \varphi^+(\theta))$ in which the oriented line $(\theta, \varphi)$ meets $P$ (or, rather, $\tilde{P}(\theta)$) is the intersection of $(\varphi^-(\theta), \varphi^+(\theta))$ with the union of two antipodal (possibly empty) angular intervals, which we denote by $T_P^+(\theta)$, $T_P^-(\theta)$, where $T_P^+(\theta)$ (resp., $T_P^-(\theta)$) consists of polar orientations of lines $\ell$ through $\ell_0(\theta) = \ell \cap \ell_0$, within $\tilde{\Pi}_\theta$, such that $\ell_0(\theta)$ lies before (resp., after) $\ell \cap P$ along $\ell$; see Figure 5. However (and this is the crucial place where we exploit the spherical arrangement $\mathcal{A}(C_H)$), since $P \in \mathcal{P}^+$ at the partition of $\mathcal{P}$ induced by $D$ and $I$ we focus only on $T_P^+(\theta)$, and rule out $T_P^-(\theta)$.

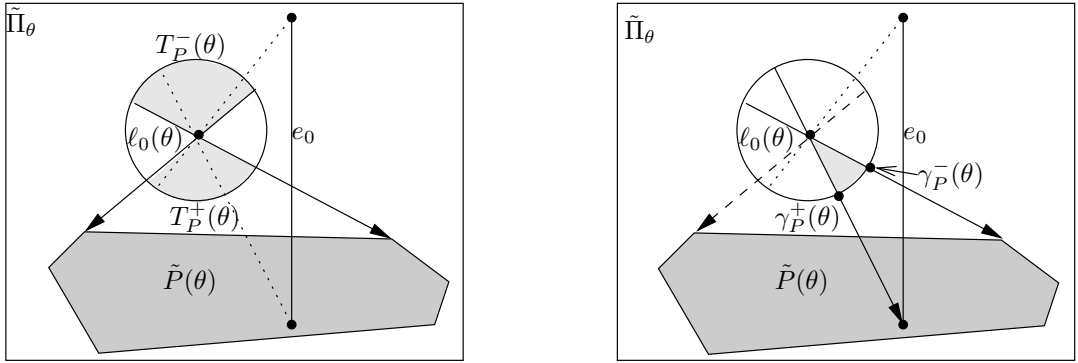

Figure 5: The polar orientations of lines in $\tilde{\Pi}_\theta$ which which pass through $\ell_0$ and stab $\tilde{P}(\theta)$ form a double wedge $T_P^+(\theta) \cup T_P^-(\theta)$ (left). The intersection of this double wedge with the range of "legal" $\varphi$-values also consists of two connected intervals, but, assuming $P \in \mathcal{P}^+$, only one of them, the depicted interval $[\gamma_P^-(\theta), \gamma_P^+(\theta)]$, survives (right).

We set $\gamma_P^-(\theta)$ (resp., $\gamma_P^+(\theta)$) to be the smallest (resp., largest) value of $\varphi$ in $T_P^+(\theta)$ (for $\varphi \in (\varphi^-(\theta), \varphi^+(\theta))$). Note that there are cases in which none of the intervals $T_P^+(\theta), T_P^-(\theta)$ survive. This would happen for example if they are empty to begin with, because $\tilde{P}(\theta)$ is empty, or if all transversals of $P$ emanating from $\ell_0(\theta)$ within $\tilde{\Pi}_\theta$ do not touch $e_0$. In any of these cases, we put $\gamma_P^-(\theta) = \varphi^+(\theta)$ and $\gamma_P^+(\theta) = \varphi^-(\theta)$. This will cause the transversal region in $\mathcal{L}[e_0, D]$ to be empty at $\theta$, see below for details.

The case where $P \in \mathcal{P}^-$ is handled in a fully symmetric manner. Note that since we assume that $\mathcal{L}[e_0, D]$ is not empty, and that $\ell_0(\theta)$ is on the same side of $\ell_{e_0}$ in $\tilde{\Pi}_\theta$, for every $\theta \in (0, \theta_0)$, then $P_0$ (the polyhedron containing $e_0$) is in $\mathcal{P}^+$ if $\ell_0(\theta)$ is to the left of $\ell_{e_0}$, and $P_0 \in \mathcal{P}^-$ if $\ell_0(\theta)$ is to the right of $\ell_{e_0}$. In both cases we have $\gamma_{P_0}^+(\theta) = \varphi^+(\theta)$ and $\gamma_{P_0}^-(\theta) = \varphi^-(\theta)$.

Note also that we do not guarantee that the great circular arc $\{\theta\} \times [\gamma_P^-(\theta), \gamma_P^+(\theta)]$ is contained in $D$. See Lemma 3.3 for a clarification of this issue. Still, since $D$ is spherically convex (and contained in a hemisphere), this arc intersects $D$ in a connected subarc.

**Proof of Lemma 3.2.**

**Lemma B.1.** *Let $P$ be a polyhedron of $\mathcal{P} \setminus \{P_0\}$. Then each algebraic subarc $\zeta$ of $\gamma_P^+$, defined by some fixed edge $e$ of $P$, is (a trace of) a regulus defined by $e_0$ and $e$. Namely, $\zeta$ represents a maximal connected set of (oriented) lines in $\mathcal{L}$, that are tangent to $P_0$ at $e_0$, and to $P$ at $e$.*

*Proof.* Indeed, we have seen that any algebraic subarc of $\gamma_P^+$ represents a connected portion of $\mathcal{L}[e_0]$ (and, therefore, $\mathcal{L}$) whose lines are tangent to $P$ at some fixed edge of it. Since each such regulus corresponds to the boundary edge of $\mathcal{T}_{\ell_0}(\{P_0, P\})$ (see Section 2), it remains to show that each breakpoint of $\gamma_P^+$, namely, the point where some algebraic arc of $\gamma_P^+$ starts or ends, corresponds to a vertex of $\mathcal{T}_{\ell_0}(\{P_0, P\})$. The transition points where $\gamma_P^+$ changes qualitatively are of the following types:

(i) $\tilde{P}(\theta)$ is a singleton point, namely, a vertex $v$ of $P$. The line $\ell^+(\theta)$ represented by $(\theta, \gamma_P^+(\theta))$ is tangent to $P$ at $v$.

(ii) $\ell^+(\theta)$ changes the edge of $P$ it is tangent to. This can happen either when $\ell^+(\theta)$ passes through a vertex of $P$ (while continuing to be tangent to $P$), or when it overlaps with a facet of $P$.

(iii) $\ell_0(\theta)$ coincides with an endpoint of an interval in $\mathcal{I}$ which lies on $\partial P$. At that moment, $\ell^+(\theta)$ overlaps a facet of $P$.

(iv) The line $\ell^+(\theta)$ passes through a vertex of $e_0$, and, in the same time, is tangent to $\tilde{P}(\theta)$ in $\tilde{\Pi}_\theta$.

As was shown in [7], at each one of the events (i)-(iv) $(\theta, \gamma_P^+(\theta))$ represents an extremal stabbing line to $\{P_0, P\}$, which completes our proof. $\qquad\square$

**Proof of Lemma 3.3.**

**Lemma B.2.** *Let $e_0$, $D$, and $P$ be as above, and let $\ell$ be an oriented line in $\mathcal{L}[e_0, D]$, with (the local) spherical coordinates $(\theta, \varphi)$. Then $\ell$ stabs $P$ if and only if*

$$\gamma_P^-(\theta) \leq \varphi \leq \gamma_P^+(\theta). \tag{6}$$

*Proof.* If $\ell_0(\theta) \notin P$, then $\ell$ stabs $P$ if and only if $\varphi$ lies in one of the two angular intervals $T_P^-(\theta)$, $T_P^+(\theta)$. (Note that the choice of $\ell \in \mathcal{L}[e_0]$ trivially ensures that $\varphi$ also lies in the interval $(\varphi^-(\theta), \varphi^+(\theta))$.) However, since $\ell \in \mathcal{L}[e_0, D]$, $\varphi$ can only belong to one of these intervals (that is, to $T_P^-(\theta)$ if $P$ belongs to the corresponding set $\mathcal{P}^-$, or to $T_P^+(\theta)$, if $P \in \mathcal{P}^+$). The claim now follows by definition.

If $\ell_0(\theta) \in P$, then $\ell$ stabs $P$, and by definition, (6) becomes $\varphi^-(\theta) \leq \varphi \leq \varphi^+(\theta)$, which holds trivially. $\qquad\square$

## C  A Recurrence for $N(n, k)$

We show the derivation of the recurrence formula for $N(n, k)$. For any line $\ell \in \mathcal{L}$, define the *depth* of $\ell$ to be the number of polyhedra of $\mathcal{P}$ which are *not stabbed* by $\ell$. For any integer $t \geq 0$, denote by $N_{\leq t}(n, k)$ the number of extremal lines in $\mathcal{L}$ whose depth is at most $t$. Since each extremal line in $\mathcal{L}$ is defined by at most three polyhedra, the standard probabilistic argument of Clarkson and Shor [10] implies that

$$N_{\leq t}(n, k) = \mathbf{E}\left\{ O\left( t^3 N\left( n_R, \frac{k}{t} \right) \right) \right\},$$

where the expectation is with respect to a random sample $\mathcal{R}$ of $\frac{k}{t}$ polyhedra of $\mathcal{P}$, and where $n_\mathcal{R}$ is the random variable equal to the number of facets in the sampled polyhedra. Clearly, the expected value of $n_R$ is $\frac{n}{t}$. To simplify the presentation, we will rewrite the Clarkson-Shor bound as

$$N_{\leq t}(n, k) = O\left( t^3 N\left( \frac{n}{t}, \frac{k}{t} \right) \right). \tag{7}$$

The argument is similar to those used in earlier works, such as in [14].

Let $r$ be a regulus of lines passing through $\ell_0$ that are tangent to $P_0$ at $e_0$, and to some other polyhedron $P$, at a fixed edge $e$ on its boundary. As is shown in Section 2, $r$ corresponds to an edge of $\mathcal{T}_{\ell_0}(\{P_0, P\})$, and there is a total of $O(nk)$ such edges, over all choices of $e_0$ and $P$. Since lines in $r$ have one degree of freedom, $r$ is represented by a connected subarc of a constant-degree curve, within the $(\theta, \varphi)$-parametric space $\mathcal{L}[e_0]$ (or, rather, within an algebraically re-parametrized version of it). To recursively bound $N(n, k)$, we distinguish between two possible cases.

(i) There is at most one cell $D \in \mathcal{A}(\mathcal{C}_H)$ such that $r$ contains an extremal stabbing line $\ell \in \mathcal{L}[e_0, D]$. Then $r$ contributes at most one unit to the sum $\sum_{D \in \mathcal{A}(\mathcal{C}_H)} R[e_0, D]$. The total contribution of all such reguli to the sum $R = \sum_{e_0, D} R[e_0, D]$ is $O(nk)$.

(ii) There are $m > 1$ cells $D \in \mathcal{A}(\mathcal{C}_H)$ such that $r$ contains an extremal stabbing line in $\mathcal{L}[e_0, D]$. Each such cell $D$ contains a connected portion $\tilde{r}$ of $r \cap D$, one of whose endpoints, call it $a$, belongs to the boundary of $D$ and the other, call it $b$, represents an extremal stabbing line in $\mathcal{L}[e_0, D]$.

Choose a positive threshold $t$. If the depth of (the line represented by) $a$ is less than $t$, we charge the appearance of $r$ in $D$ to $a$. Since $a$ is contained in the boundary of $D$, there is a plane $h \in H$ such that $a$ represents a line in the space $\mathcal{L}_h$ of lines which pass through $\ell_0$ and are parallel to $h$. Moreover, since $r$ is a common tangent to $P_0$ and $P$, and lines in $\mathcal{L}_h$ have two degrees of freedom, $a$ represents an extremal line at depth $\leq t$, within $\mathcal{L}_h$. By Lemma 2.2, the number of extremal lines at depth 0 within $\mathcal{L}_h$ is $O(n\beta_4(k))$. By applying the Clarkson-Shor argument [10], as above, and using the fact that an extremal stabbing line in $\mathcal{L}_h$ is defined by at most two polyhedra, we conclude that the number of such extremal lines at depth $\leq t$, within $\mathcal{L}_h$, is $O(t^2 \cdot \mathbf{E}\{n_R \beta_4(k/t)\})$, where $n_R$ is the number of facets in a random sample of $k/t$ polyhedra of $\mathcal{P}$. Since $\mathbf{E}\{n_R\} = n/t$, the bound is $O(tn\beta_4(k/t))$. Since $|H| = O(k)$, the overall number of lines $a$, as above, of depth smaller than $t$, is $O(tnk\beta_4(k/t)) = O(tnk\beta_4(k))$. Moreover, any such line $a$ is charged at most a constant number of times, over all possible choices of $e_0$ and $D$.

If the depth of (the line represented by) $a$ is at least $t$, we walk from $b$ to $a$ along $\tilde{r}$, and collect at least $t$ extremal lines at depth $\leq t$ contained in $\mathcal{L}[e_0, D]$. We charge the appearance of $r$ in $D$ to those lines. Clearly, each line is charged in this manner only a constant number of times, over all choices of $e_0$ and $D$. Using (7), and taking case (i) also into account, we get

$$R(n, k) = O\left(nk + tnk\beta_4(k) + \frac{1}{t} \cdot t^3 N\left(\frac{n}{t}, \frac{k}{t}\right)\right).$$

Combining this with the relation $N(n, k) = O(R(n, k)\beta_4(k))$. that follows from (4), we finally get a recurrence for $N(n, k)$:

$$N(n, k) \leq ctnk\beta_4^2(k) + ct^2 N\left(\frac{n}{t}, \frac{k}{t}\right)\beta_4(k), \tag{8}$$

for an appropriate constant $c$.

The recurrence terminates when $k$ becomes smaller than $t$. In this case, we use the fact, which was proved in [7], that the complexity of the arrangement, within $\mathcal{L}$, of the upper and lower tangency surfaces $\sigma_P^+, \sigma_P^-$ of all polyhedra, is bounded by $O(nk^2) = O(nt^2)$, which thus also serves as an upper bound for $N(n, k)$.

Using standard analysis, such as that in [4, 14], it follows that the solution of the recurrence in (8) is $N(n, k) = O(nk^{1+\varepsilon})$, for any $\varepsilon > 0$, where the constant of proportionality depends on $\varepsilon$ (the choice of $\varepsilon$ affects the choice of $t$). This completes the proof of the upper bound in Theorem 3.1. [7]

---

[7]The fact that the bound that we derive is linear in $n$ justifies the bound (7). Although this deserves a more

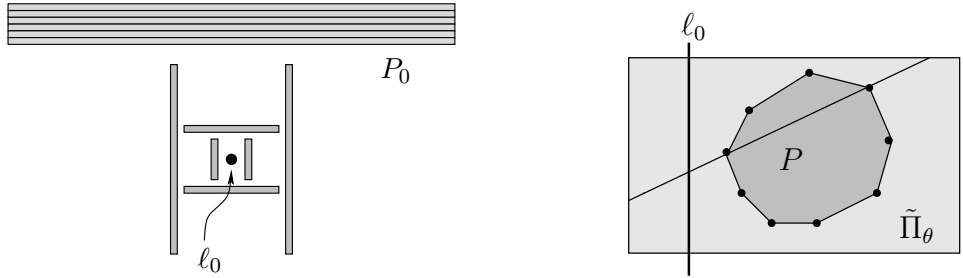

Figure 6: (Left) A lower-bound construction, viewed from above. (The actual gaps between the pairs of parallel "plates" around $\ell_0$, relative to the width of the plates, are much smaller than depicted.) (Right) If $\mathcal{P}$ consists of a single polygon, which is coplanar with $e_0$, then every pair of distinct vertices of $P$ define an extremal stabbing line passing through them.

## D  Proof of the Lower Bound of Theorem 3.1

A lower bound construction for the complexity of $\mathcal{T}_{\ell_0}(\mathcal{P})$ is depicted in Figure 6 (left). We use $k$ pairs of polyhedra, each of which is a thin and long vertical plate, whose vertical edges and facets are all parallel to $\ell_0$. The polyhedra in each pair are parallel to each other, and situated symmetrically around $\ell_0$. In addition, we include in $\mathcal{P}$ one drum-like polyhedral prism $P_0$, whose axis is horizontal (i.e., orthogonal to $\ell_0$), and which has $n \gg k$ long and narrow facets. Overall, we have $\Theta(k)$ polyhedra with a total of $\Theta(n)$ facets.

The polyhedra in $\mathcal{P} \setminus \{P_0\}$ are arranged so that, as we rotate a line of $\mathcal{L}$ around $\ell_0$, by varying its $\theta$-component, while keeping its $z$- and $\varphi$-components within some reasonable range, we obtain $\Theta(k)$ distinct geometric permutations of the polyhedra in $\mathcal{P} \setminus \{P_0\}$ (see Figure 6, and [9, 11] for similar constructions). With an appropriate choice of the layout, we can construct, for each of these permutations and for each edge $e$ of $P_0$, a line in $\mathcal{L}$ that realizes the permutation, and is tangent to $P_0$ at $e$ and to two of the other polyhedra at two respective vertical edges. We thus obtain a lower bound of $\Omega(nk)$ on the complexity of $\mathcal{T}_{\ell_0}(\mathcal{P})$. (The construction is not in general position, but can be transformed into general position by a small perturbation of its polyhedra.)

This completes the proof of Theorem 3.1.

**Remark:** For the bound of Theorem 3.1 to hold, we have to assume general position of $\ell_0$ and the polyhedra of $\mathcal{P}$, for otherwise it is easy to construct scenarios with $\Omega(n^2)$ extremal transversals, as depicted in the Figure 6 (right).

## E  Proof of Theorem 3.5

**Theorem E.1.** *Let $\mathcal{P}$ be a collection of $k$ convex polyhedra in $\mathbb{R}^3$ with a total of $n$ facets. Then the set $\mathcal{T}(\mathcal{P})$ of line transversals of $\mathcal{P}$ has complexity $O(n^2 k^{1+\varepsilon})$, for any $\varepsilon > 0$.*

*Proof.* For the sake of simplicity, we only bound the number of extremal stabbing lines in $\mathcal{T}(\mathcal{P})$,

---

formal argument, we omit the details, which are similar to those in earlier studies [14, 26].

which correspond to boundary vertices of the stabbing region, within an appropriate parametric space of lines; see [24] and Section 2 for a standard justification.

Let then $e_0$ be a fixed edge of the boundary of some polyhedron $P_0 \in \mathcal{P}$, let $\ell_0$ be the line which contains $e_0$, and let $\mathcal{L}$ be the space of lines passing through $\ell_0$. Clearly, any extremal stabbing line to $\mathcal{P}$ that is tangent to $P_0$ at the relative interior of $e_0$ is also an extremal stabbing line $\ell$ to $\mathcal{P} \setminus \{P_0\}$ in $\mathcal{L}$, unless it passes through an endpoint of $e_0$, or it overlaps a facet of $P_0$ incident to $e_0$. Now using Theorem 3.1, and summing over $O(n)$ possible choices of $e_0$, we immediately derive an $O(n^2 k^{1+\varepsilon})$ upper bound, for any $\varepsilon > 0$, on the number of extremal stabbing lines that are tangent to at least one polyhedron at the relative interior of one of its edges, and not overlapping any of its facets.

To complete our proof, we next show an upper bound of $O(n^2 k)$ on the number of remaining extremal stabbing lines, each of which, assuming general position of $\mathcal{P} \cup \{\ell_0\}$, is defined by two or fewer polyhedra of $\mathcal{P}$. For the sake of simplicity, we derive the above bound only for extremal stabbing lines defined by pairs of polyhedra. Fix any pair of distinct polyhedra $P, Q \in \mathcal{P}$, with $n_P$ and $n_Q$ facets, respectively. As shown by Brönnimann et al. [7], the number of such extremal stabbing lines defined by $P$ and $Q$ is $O((n_P + n_Q)^2)$. Summing over all choices of $P$ and $Q$, the number of extremal stabbing lines of this type is

$$O\left(\sum_{P,Q \in \mathcal{P}, P \neq Q} (n_P + n_Q)^2\right) = O\left(\sum_{P,Q \in \mathcal{P}, P \neq Q} (n_P^2 + n_Q^2)\right) = O(n^2 k) .$$

□

**Remarks:** (a) Unlike the restricted case analyzed in Section 3, here we do not have a matching lower bound. The best lower bound is the trivial $\Omega(n^2)$ bound; see [7]. A different lower bound of $\Omega(nk^2)$ can be shown by slightly modifying the lower-bound construction of Theorem 3.1. Closing the gap between the resulting lower bound $\Omega(n^2 + nk^2)$ and the upper bound in Theorem 3.5 remains a major challenging open problem.

(b) Unlike the situation in Section 3 (see the remark at the end of that section), Theorem 3.5 continues to hold also without the general position assumption. This can be shown, e.g., using small random perturbation of the input polyhedra.

## F  Proof of Theorem 3.6

**Theorem F.1.** *Let $\mathcal{P}$ be a set of $k$ convex polyhedra with a total of $n$ facets, and let $\ell_0$ be a fixed line (all in general position). Then one can compute the stabbing region $\mathcal{T}_{\ell_0}(\mathcal{P})$ of $\mathcal{P}$, within the set $\mathcal{L}$ of lines passing through $\ell_0$, in $O(nk^{1+\varepsilon} \log n)$ randomized expected time, for any $\varepsilon > 0$.*

*Proof.* Without any loss of generality, we assume, as above, that $\ell_0$ is the $z$-axis. Recall that for each $P \in \mathcal{P}$, the graphs of $\sigma_P^+$ and $\sigma_P^-$, respectively, are $(\theta, \varphi)$-monotone surfaces consist of monotone semi-algebraic surface patches, each of which is a graph of a partially-defined function, representing lower (resp., upper) tangents to $P$ at a fixed edge of its lower (resp., upper) boundary. For each edge $e$ of $P$, we denote its corresponding tangency function (and its graph) by $\sigma_e$. The domain of $\sigma_e$ is the region $D_e \cup \tilde{D}_e$, where $\tilde{D}_e$ is obtained from $D_e$ by mapping[8] $(\theta, \varphi)$ to $(\pi + \theta, \pi - \varphi)$, and

$$D_e = \{(\theta, \varphi) \mid \theta_e^- \leq \theta \leq \theta_e^+,\ \tau_{f^-}(\theta) \leq \varphi \leq \tau_{f^+}(\theta)\}, \tag{9}$$

where $\theta_e^-$ and $\theta_e^+$ are the $\theta$-coordinates of the two (clockwise and counterclockwise) endpoints of $e$, and where $f^-$ and $f^+$ are the two facets of $P$ incident to $e$, and $\tau_{f^-}$ (resp., $\tau_{f^+}$) is the locus of all orientations $(\theta, \varphi)$ of lines parallel to $f^-$ (resp., $f^+$); since, as is easily checked, the sets $\tau_{f^-}$, $\tau_{f^+}$ are $\theta$-monotone, the functional notation (9), as well as the determination of which of the adjacent facets is $f^-$ and which is $f^+$, are well defined. If $e$ is a silhouette edge of $P$, i.e., it admits a supporting plane of $P$ parallel to $\ell_0$, then we consider it as a pair of identical copies, one of which belongs to the upper portion of $\partial P$, and the other to the lower portion of $\partial P$. In each of these cases, one of $\tau_{f^+}$ or $\tau_{f^-}$, as appropriate, is set to $+\infty$ or $-\infty$, respectively.

As a preparatory step, we use the algorithm of Theorem 2.1 to construct $\mathcal{T}_{\ell_0}(\{P, Q\})$, for each pair of polyhedra $P, Q \in \mathcal{P}$, in a total of $O(nk \log n)$ time. In particular, for each edge $e$ of the upper (resp., lower) boundary of some polyhedron $P$, and for each $Q \in \mathcal{P} \setminus \{P\}$, the above algorithm computes the intersection of $\sigma_Q^+$ and of $\sigma_Q^-$ with the (two-dimensional) patch $\sigma_e$ representing upper (resp., lower) tangents to $P$ at $e$ (recall that silhouette edges are treated as both upper and lower edges).

The algorithm for constructing $\mathcal{T}_{\ell_0}(\mathcal{P})$ proceeds as follows. First, we choose a fixed random permutation $\pi$ of the polyhedra of $\mathcal{P}$. Next, we fix a polyhedron $P \in \mathcal{P}$, and a boundary edge $e$ of $P$. Recall that $\sigma_e$ consists of two connected patches of a constant-degree algebraic surface (in an appropriate re-parametrization); one patch is defined over the domain $D_e$ given in (9), and the other is defined over the symmetric domain $\tilde{D}_e$. Without loss of generality, we consider the portion of $\sigma_e$ defined over $D_e$. To obtain $\mathcal{T}_{\ell_0}(\mathcal{P})$, it is sufficient to compute $\sigma_e \cap \mathcal{T}_{\ell_0}(\mathcal{P} \setminus \{P\})$, for all possible choices of $P$ and $e$ as above; the union of all these 2-dimensional patches constitutes the boundary of $\mathcal{T}_{\ell_0}(\mathcal{P})$. We follow a standard randomized divide-and-conquer approach; see [27]. Let $\mathcal{P}_R$ be the set of the first $\lfloor \frac{k-1}{2} \rfloor$ elements of $\mathcal{P} \setminus \{P\}$ in the permutation $\pi$, and let $\mathcal{P}_B$ be the set of the $\lceil \frac{k-1}{2} \rceil$ remaining elements of $\mathcal{P} \setminus \{P\}$. We start by recursively computing the "red region" $\mathcal{R} := \sigma_e \cap \mathcal{T}_{\ell_0}(\mathcal{P}_R)$ and the "blue region" $\mathcal{B} := \sigma_e \cap \mathcal{T}_{\ell_0}(\mathcal{P}_B)$ within $\sigma_e$. Then $\sigma_e \cap \mathcal{T}_{\ell_0}(\mathcal{P} \setminus \{P\}) = \mathcal{R} \cap \mathcal{B}$. We denote this intersection by $\mathcal{G}$, and refer to it as the "green region". Let $n_R$, $n_B$, and $n_G$ denote the total number of vertices and edges on the boundary of $\mathcal{R}$, $\mathcal{B}$, and $\mathcal{G}$, respectively. Then $\mathcal{G}$ can be computed in $O((n_R + n_B + n_G) \log(n_R + n_B + n_G))$ time, using (an appropriate variant of) a standard planar sweep algorithm; see, e.g., [27]. We repeat this procedure for all possible choices of $P$ and $e$. The above algorithm has expected running time $O(N \log N \log k)$, where $N$ is the overall expected number of vertices and edges which are constructed by the algorithm at any recursive step, over all choices of $P$ and $e$.

We next establish an upper bound on $N$. Note that any edge of $\mathcal{T}_{\ell_0}(\mathcal{P})$ which does not have any incident vertex, and which appears in one of the regions constructed by the algorithm, appears in $\mathcal{T}_{\ell_0}(\{P, Q\})$, for some pair of polyhedra $P, Q \in \mathcal{P}$, so, according to Theorem 2.1, the overall number of such edges is $O(nk)$. The appearance of any other edge can be charged to that of an incident vertex, so that each vertex is charged at most a constant number of times. Therefore, it remains to bound the expected number of vertices which appear in some region $\mathcal{G}$, throughout the execution of the algorithm (over all $P$ and $e$). Clearly, the overall number of vertices of the curves $\sigma_e \cap \sigma_Q^+, \sigma_e \cap \sigma_Q^-$, over all choices of $P$, $e \in \partial P$, and $Q \in \mathcal{P} \setminus \{P\}$, is $O(nk)$, since those vertices appear in the respective regions $\mathcal{T}_{\ell_0}(\{P, Q\})$. Hence, it suffices to bound the expected number of vertices defined by triples of polyhedra. For a fixed choice of $P$ and $e$, any such vertex $v$ belongs to the intersection of $\sigma_e \cap \sigma_Q^+$ (or $\sigma_e \cap \sigma_Q^-$), and $\sigma_e \cap \sigma_R^+$ (or $\sigma_e \cap \sigma_R^-$), for some pair of distinct polyhedra $Q, R \in \mathcal{P} \setminus \{P\}$. Let $\mathcal{K}_v$ denote the set of polyhedra which are not stabbed by the extremal line $\ell_v$ corresponding to $v$. Thus, $|\mathcal{K}_v|$ is equal to the depth of $\ell_v$. As is easily verified, a necessary condition for $v$ to appear in $\mathcal{G}$, at some recursive step, is that no polyhedron of $\mathcal{K}_v$ appears between $Q$ and $R$ in $\pi$. We say that $v$ has depth $t$ if $\ell_v$ has depth $t$. Restricting $\pi$ to the $t + 2$ polyhedra $Q, R$, and the $t$ polyhedra that $\ell_v$ misses, it follows that the probability that $v$ is constructed by the algorithm, when processing $\sigma_e$, is at most

$$\frac{2(t+1)!}{(t+2)!} = \frac{2}{t+2}.$$

---

[8]That is, if an oriented line $\ell \in \mathcal{L}$ has orientation $(\theta, \varphi) \in D_e$ then the oppositely-oriented copy of $\ell$ has orientation $(\theta + \pi, \pi - \varphi) \in \tilde{D}_e$.

Note that $v$ may also arise when processing the respective edges of $Q$ and of $R$ which define $v$, so we actually need to multiply the corresponding probabilities and expectations by 3.

Let $N_{\le t}$ denote the number of extremal lines at depth at most $t$, summed over all $P$ and $e$. Then the overall number of extremal lines at depth $t$, for $0 \le t \le k - 3$, is $N_{\le t} - N_{\le t-1}$, where $N_{\le -1} = 0$, and the expected number of vertices at depth $t$, which appear in some region $\mathcal{G}$ throughout the algorithm, again over all $P$ and $e$, is at most

$$\frac{6(N_{\le t} - N_{\le t-1})}{t + 2}.$$

Summing over all depths $0 \le t \le k - 3$, we get

$$N \le 6 \sum_{0 \le t \le k-3} \frac{(N_{\le t} - N_{\le t-1})}{t + 2} + O(nk). \tag{10}$$

Rearranging the sum, we get

$$N \le 6 \sum_{0 \le t \le k-4} N_{\le t} \left( \frac{1}{t+2} - \frac{1}{t+3} \right) + \frac{6}{k-1} N_{\le k-3} + O(nk) \tag{11}$$

$$= 6 \sum_{0 \le t \le k-4} \frac{N_{\le t}}{(t+2)(t+3)} + \frac{6}{k-1} N_{\le k-3} + O(nk).$$

By Theorem 3.1, for any $\varepsilon > 0$, $N_{\le 0} = O(nk^{1+\varepsilon})$. Hence, by the Clarkson-Shor probabilistic argument [10], using (7), $N_{\le t} = O(tnk^{1+\varepsilon})$, for all $1 \le t \le k - 3$; see [4] for a similar argument. Plugging this into (11), we obtain that $N = O(nk^{1+\varepsilon})$, for any $\varepsilon > 0$. (The substitution yields a harmonic series which adds a factor $O(\log k)$ to the bound, but this factor is "swallowed" by $k^{1+\varepsilon}$, by slightly increasing $\varepsilon$, still keeping it arbitrary small.) Our algorithm thus runs in $O(nk^{1+\varepsilon} \log n)$ randomized expected time, for any $\varepsilon > 0$. (Again, the $O(\log k)$ factor yielded by the divide-and-conquer process is subsumed in the factor $k^{1+\varepsilon}$.) □

## G  Extensions

Here we prove the theorems of Section 4.

## G.1    $\ell_0$ is disjoint from all polyhedra in $\mathcal{P}$

**Theorem G.1.** *Let $\mathcal{P}$ be a set of $k$ pairwise disjoint convex polyhedra with a total of $n$ facets, and let $\ell_0$ be a fixed line disjoint from all polyhedra of $\mathcal{P}$ (all in general position). Then the complexity of the stabbing region $\mathcal{T}_{\ell_0}(\mathcal{P})$ of $\mathcal{P}$ in the space $\mathcal{L}$ of lines passing through $\ell_0$ is $O((nk+k^3)\beta_4(k))$. Moreover, one can compute $\mathcal{T}_{\ell_0}(\mathcal{P})$ in $O((nk+k^3)(\log n + \alpha(k)\log k))$ deterministic time.*

*Proof.* To bound the number of extremal stabbing lines in $\mathcal{L}$, we return to the proof of Theorem 3.1. Recall that, in the current context where $\ell_0$ is disjoint from all the polyhedra of $\mathcal{P}$, $H$ is a collection of $k$ planes, each separating a polyhedron $P \in \mathcal{P}$ from $\ell_0$. Clearly, we can choose $H$ such that all of its planes contain $\ell_0$.

Let $e_0$ be a boundary edge of some polyhedron $P_0 \in \mathcal{P}$, and let $D$ be a cell in the arrangement $\mathcal{A}(\mathcal{C}_H)$, as defined above. Recall that $R[e_0, D]$ is defined to be the number of arcs that actually show up on the boundary of the sandwich region (3), at points which represent lines in $\mathcal{L}[e_0, D]$. Then $R[e_0, D]$ is bounded by the number of reguli of lines (in $\mathcal{L}$) that are tangent to $P_0$ at $e_0$, and to another polyhedron $P \in \mathcal{P}$ at some fixed boundary edge $e$, and contain a line with direction in $D$. (Here we relax the requirement that such a regulus contains a *stabbing* line with direction in $D$.) For each such regulus $r$, we denote by $r^*$ the locus, on $SS^2$, of the orientations of the lines in $r$. Recall that we only consider those portions of the reguli consisting of lines that pass through the respective edges $e_0$ and $e$.

Recall that the number of extremal stabbing lines (in $\mathcal{L}$) is

$$O\left(\sum_{P_0 \in \mathcal{P}, e_0 \in \partial P} \sum_{D \in \mathcal{A}(\mathcal{C}_H)} R[e_0, D]\beta_4(k)\right).$$

As noted in Section 2, there is a total of $O(nk)$ reguli, over all choices of $e_0$. Hence, the sum

$$\sum_{P_0 \in \mathcal{P}, e_0 \in \partial P} \sum_{D \in \mathcal{A}(\mathcal{C}_H)} R[e_0, D] \tag{12}$$

is bounded by $O(nk)$ plus a constant times the number of crossings of reguli $r$ and boundaries of cells $D \in \mathcal{A}(\mathcal{C}_H)$. Any such intersection corresponds to a distinct (oriented) line, *contained* in some plane $h \in H$ and tangent to a pair of polyhedra of $\mathcal{P}$. (By the choice of $H$, a line in $\mathcal{L}$ is parallel to a plane $h \in H$ if and only if it lies in $H$.) Since the polyhedra of $\mathcal{P}$ are pairwise disjoint, the number of such lines is $O(k^2)$, for a fixed $h \in H$. Since $|H| = k$, we get that the sum in (12) is $O(nk + k^3)$. Hence, the number of extremal stabbing lines in $\mathcal{L}$, and thus the complexity of the stabbing region $\mathcal{T}_{\ell_0}(\mathcal{P})$, is $O((nk+k^3)\beta_4(k))$. Note that this bound constitutes an improvement over Theorem 3.1 when $k = O(\sqrt{n})$.

To compute the extremal stabbing lines in $\mathcal{L}$, we first compute the set of all reguli $r$, in $O(nk \log n)$ overall time, using the algorithm of Theorem 2.1. Then, for each such regulus $r$, we trace $r^*$ through the arrangement $\mathcal{A}(\mathcal{C}_H)$, in $O((nk+k^3)\log k)$ total time. To do so, we prepare $\mathcal{A}(C_H)$ for point location queries, and then locate, for each regulus $r$, the cell of $\mathcal{A}(C_H)$ that contains the orientation of an "endpoint" line of $r$. This takes $O(nk \log k)$ time. Then, for each plane $h \in H$, we intersect the polyhedra of $P$ with $h$, and find the $O(k^2)$ common tangent lines (in $h$) to pairs of these intersections. This takes $O(n + k^2 \log n)$ time for each $h$, for a total of $O(nk + k^3 \log n)$ time. Then, we take each such common tangent, and locate the arc of $\mathcal{A}(C_H)$ (on $C_h$) which contains its orientation. Combining the outputs of this and the preceding step, we obtain the sets $R[e_0, D]$, for all edges $e_0$ and cells $D$, in overall $O((nk+k^3)\log n)$ time (see the preceding analysis for justification of this procedure). Finally, we compute, for each choice of $P_0$, $e_0$, and $D$, the vertices of the sandwich region, defined in the proof of Lemma 3.1, in $O(R[e_0, D]\beta_3(k)\log k) = O(R[e_0, D]\alpha(k)\log k)$ time, using the algorithm of Hershberger [16]; see also [27, Theorem 6.5]. Summing over all choices of $P_0$, $e_0$, and $D \in \mathcal{A}(\mathcal{C}_H)$, the total construction cost is $O((nk + k^3)\alpha(k)\log k)$ time. We can thus compute the extremal stabbing lines in $\mathcal{L}$ in $O((nk + k^3)(\log n + \alpha(k)\log k))$ total time.

Computing the other features of $\mathcal{T}_{\ell_0}(\mathcal{P})$, such as edges without incident vertices, can also be done within this time bound (see Section 2 for details). This completes the proof of the theorem. □

## G.2 The polyhedra of $\mathcal{P}$ are unbounded in a direction parallel to $\ell_0$

For this subsection, we represent lines in $\mathcal{L}$ as in Section 2, and establish the following result, which holds for more general collections of convex polyhedra.

**Theorem G.2.** *Let $\mathcal{P}$ be a collection of $k$ pairwise disjoint convex polyhedra in $\mathbb{R}^3$, having a total of $n$ facets, and let $\ell_0$ be a fixed line. The number of vertices of the upper envelope $E_U$ of the partially-defined functions $\sigma_P^-$, for $P \in \mathcal{P}$ (as defined in Section 2), is $O(nk\beta_4(k))$. Moreover, there is a deterministic algorithm which computes $E_U$ in $O(nk(\log n + \alpha(k)\log k))$ time.*

*Proof.* As in Section 2, we assume that $\ell_0$ is the $z$-axis. Fix a polyhedron $P$ of $\mathcal{P}$, and consider the graph of the function $z = \sigma_P^-(\theta, \varphi)$, which represents lower tangents to $P$ in $\mathcal{L}$. We denote this graph also as $\sigma_P^-$, and assume, without loss of generality, that $\sigma_P^-$ is not empty. For each $Q \in \mathcal{P} \setminus \{P\}$, let $\sigma_{PQ}^-$ denote the (relatively open) portion of $\sigma_P^-$ that lies *below* $\sigma_Q^-$ (in the $z$-direction); that is, $\sigma_Q^-$ is higher than $\sigma_P^-$ over this portion, and thus "prevents" $\sigma_P^-$ from attaining $E_U$ at these points. Hence, the portion of $\sigma_P^-$ that appears on $E_U$ is the complement of the union of the regions $\sigma_{PQ}^-$, for $Q \in \mathcal{P} \setminus \{P\}$.

We fix $P$ and $Q$, and study in more detail the structure of $\sigma_{PQ}^-$. Fix some value $\theta_0$ of $\theta$, and consider the cross sections of $\sigma_P^-$ and $\sigma_Q^-$ at $\theta_0$. Any line $\ell$ with $\theta = \theta_0$ lies in the plane $\Pi_\theta$. Any point on the intersection curve $\gamma_{PQ} = \sigma_P^- \cap \sigma_Q^-$, with $\theta = \theta_0$, corresponds to a line in $\Pi_{\theta_0}$ which is a common *lower* tangent to $P(\theta_0)$ and $Q(\theta_0)$. Since these two convex polygons are disjoint, they have at most two common lower tangents.

Moreover, the only way in which they can have two common lower tangents is when one of $P(\theta_0)$, $Q(\theta_0)$ lies "fully above" the other. Formally, this happens if and only if (i) the vertical projections of $P(\theta_0)$ and $Q(\theta_0)$ (on the axis of $\Pi_{\theta_0}$ orthogonal to $\ell_0$) are nested within each other, and (ii) the polygon with the larger projection, say, $Q(\theta_0)$, lies above the other polygon $P(\theta_0)$. We then say that $Q$ *overshadows* $P$ (at $\theta_0$), and denote this as $P \prec Q$ (at $\theta_0$). Clearly, for any fixed $\theta_0$, this is a partial (strict) order; see Figure 7 for an illustration.

Suppose then that $P \prec Q$ at $\theta_0$. Then, as noted, $\gamma_{PQ}$ meets $\theta = \theta_0$ at two points $(\theta_0, \varphi_1)$, $(\theta_0, \varphi_2)$, with $\varphi_1 < \varphi_2$. Moreover, as is easily checked, $\sigma_{PQ}^-$ is the portion of $\sigma_P$ (strictly) between $\varphi_1$ and $\varphi_2$. This is a bad scenario, because then $\sigma_{PQ}^-$ is (the cross-section at $\theta = \theta_0$ of) a "strip", and the union of such "strips" can have quadratic complexity; see Figure 8 for an illustration.

Before proceeding to address this issue, we note that all the other cases are "good": If neither of $P(\theta_0)$, $Q(\theta_0)$ overshadows the other, then they have at most one common lower tangent line, and then $\sigma_{PQ}^-$, at $\theta_0$, is either the portion $\varphi > \varphi_0$ or the portion $\varphi < \varphi_0$, where $(\theta_0, \varphi_0)$ is the intersection point of $\gamma_{PQ}$ with $\theta = \theta_0$ (if it exists at all; if not, $\sigma^-_{PQ}$ is either empty at $\theta = \theta_0$ or consists of the entire range of $\varphi$). If, on the other hand, $Q \prec P$ at $\theta_0$ then, arguing as above, $\sigma^-_{PQ}$ is the portion of $\sigma^-_P$ (strictly) above $\varphi_2$ or below $\varphi_1$. In either case, the union of these good regions $\sigma^-_{PQ}$ is the union of "halfspaces", where each halfspace consists of those points whose $\theta\varphi$-projections lie above some $\theta$-monotone curve (*upper halfspaces*), or below such a curve (*lower halfspaces*); here "above" and "below" are with respect to the $\varphi$-direction. Thus, for the "good" interactions involving $P$, the complement of the union is a (2-dimensional) sandwich region between the upper envelope of the (curves bounding the) lower halfspaces, and the lower envelope of the (curves bounding the) upper halfspaces. This property will be crucial for the analysis of the number of vertices of $E_U$. Note that, when $Q \prec P$, $\sigma^-_{PQ}$ is the union of two halfspaces, so we may have up to $2(k-1)$ halfspaces which form the sandwich region.

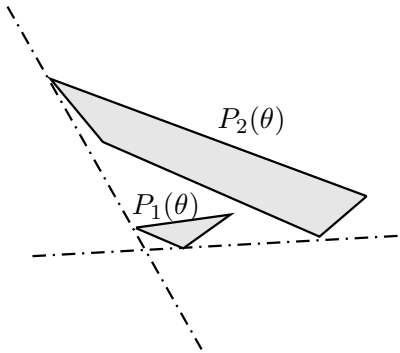

Figure 7: Two polygons $P_1(\theta)$ and $P_2(\theta)$ can have two common lower tangents only when one of them, $P_2$, overshadows $P_1$.

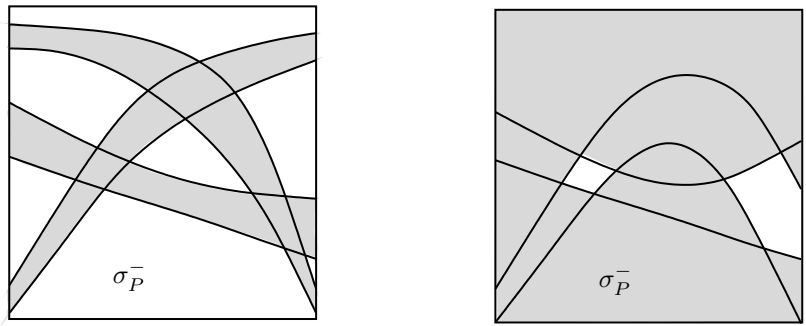

Figure 8: In the bad scenario (left), to be avoided, the regions $\sigma_{PQ}^-$ may appear as "strips" within $\sigma_P^-$, and their union may have quadratic complexity. In the good scenario (right), we only consider regions $\sigma_{PQ}^-$ that appear as "halfplanes" (or as pairs of disjoint halfplanes) within $\sigma_P^-$, in which case the complement of their union (shown unshaded) behaves like a sandwich region and has nearly linear complexity.

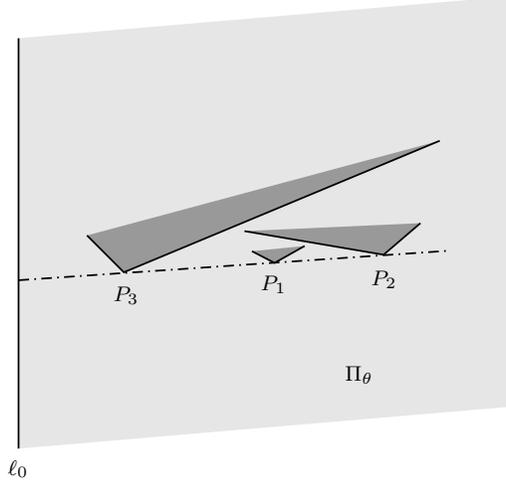

Figure 9: A critical line tangent to three polyhedra of $\mathcal{P}$ from below. Here $P_1 \prec P_2 \prec P_3$.

Our strategy is thus as follows. Let $v$ be a vertex of the upper envelope $E_U$, incident to three surfaces, which we denote as $\sigma^-_P, \sigma^-_Q, \sigma^-_R$, corresponding to three respective polyhedra $P, Q, R \in \mathcal{P}$. We claim that there exists at least one surface, say $\sigma^-_P$, such that neither of the relations $P \prec Q$, $P \prec R$ holds (at the $\theta$-coordinate of $v$). Indeed, if this does not hold for $\sigma^-_P$, then, say, $P \prec Q$. If $Q \prec R$ then, by transitivity, we also have $P \prec R$, so $R$ has the desired property (using the anti-symmetry of $\prec$). If, on the other hand, $Q \not\prec R$, then $Q$ has the desired property. See Figure 9 for an illustration.

Suppose, without loss of generality, that $P$ has this property; that is, neither of the relations $P \prec Q$, $P \prec R$ holds. By the preceding observations, both $\sigma^-_{PQ}$ and $\sigma^-_{PR}$ are good at (the vicinity of) $v$, so $v$ appears as a vertex of the sandwich region of the good regions $\sigma^-_{PS}$, for $S \in \mathcal{P} \setminus \{P\}$.

It therefore suffices to bound the overall complexity of the sandwich regions of the good portions $\sigma^-_{PQ}$, within the respective surfaces $\sigma^-_P$.

Let us study in detail the structure of a good portion $\sigma^-_{PQ}$ within $\sigma^-_P$. It is a "partially defined" halfspace, bounded by two extreme $\theta$-values $\theta_1 < \theta_2$, and lying above or below a $\theta$-monotone portion of the curve $\gamma_{PQ}$. If $Q \not\prec P$ then $\gamma_{PQ}$ itself is $\theta$-monotone. If $Q \prec P$, $\gamma_{PQ}$ has two $\theta$-monotone arcs, and, as already mentioned, we consider the portions of $\sigma^-_{PQ}$ that lie above the top arc and below the bottom arc, respectively, as two separate halfspaces.

At each of the delimiting orientations $\theta_1$, $\theta_2$, the overshadowing relation between $P(\theta)$ and $Q(\theta)$ starts or stops holding. When this happens, the vertical projections of $P(\theta)$ and $Q(\theta)$ have a common endpoint. Alternatively, if we project the polyhedra of $\mathcal{P}$ onto some plane $h_0$ orthogonal to $\ell_0$, then $\theta_1$ and $\theta_2$ are the $\theta$-coordinates of intersection points between the projected *silhouettes* of $P$ and $Q$. As is easily checked, the total number of such intersection points, over all pairs $P$, $Q$, is only $O(nk)$. (That is, in an arrangement of $k$ convex polygons in the plane, with a total of $n$ edges, there are at most $nk$ points where a pair of polygon boundaries cross each other.)

With an appropriate re-parametrization of $\theta$ and $\varphi$, each curve $\gamma_{PQ}$ is piecewise algebraic. Each piece corresponds to a regulus of lower tangent lines of $P$, $Q$ which touch $P$, $Q$ at two fixed respective edges $e, e'$. A breakpoint of $\gamma_{PQ}$ corresponds to situations where one of the incident edges $e, e'$ changes. As already argued in the proof of Theorem 3.1 (see also Theorem 2.1), the number of such breakpoints is $O(nk)$.

To recap, on each surface $\sigma_P^-$, for $P \in P$, we need to bound the complexity of a sandwich region between the lower envelope of up to $k-1$ $\theta$-monotone curves and the upper envelope of up to $k-1$ other $\theta$-monotone curves. Each curve consists of several arcs, each representing tangency with some fixed pair of edges of two of the polyhedra, and the total number of these arcs is $O(nk)$, over all surfaces $\sigma_P^-$.

We next observe that any pair of these arcs, say, $\gamma' \subset \gamma_{PQ}$, $\gamma'' \subset \gamma_{PR}$, on the same surface $\sigma_P$, intersect at most twice. Indeed, any such intersection point $p$ represents a line that passes through four lines, namely, $\ell_0$, the line containing the edge $e$ of $P$ touched by the line $\ell_p$ represented by $p$, and the lines containing the edges of $Q$ and $R$, corresponding to $\gamma', \gamma''$, respectively, touched by $\ell_p$. Since at most two lines can touch four distinct lines (not all lying on a common regulus; see [29]), the claim follows.

It therefore follows, using similar arguments to those given above, that the overall complexity of all the sandwich regions, over all surfaces $\sigma_P$, is $O(n\lambda_4(k)) = O(nk\beta_4(k))$. This completes the proof of the upper bound of the theorem.

The above analysis immediately implies the existence of a deterministic algorithm for computing the vertices of $\mathcal{T}_{\ell_0}(\mathcal{P})$ in $O(nk(\log n + \alpha(k)\log k))$ time. Indeed, we can obtain the curves $\gamma_{PQ}$, for every pair of distinct polyhedra $P, Q \in \mathcal{P}$, using the algorithm of Theorem 2.1. We then compute the good portion of each $\gamma_{PQ}$, for each pair of distinct polyhedra $P, Q \in \mathcal{P}$. Finally, we compute, for each $P \in \mathcal{P}$, the corresponding sandwich region bounded by the good portions of the curves $\gamma_{PQ}$, for $Q \in \mathcal{P} \setminus \{P\}$. As can be easily verified, the first two steps take $O(nk \log n)$ time, and the last step takes a total of $O(nk\alpha(k)\log k + nk\log n)$ time, using Hershberger's algorithm [16], where the second term is the cost of sorting arc endpoints, for each sandwich region separately. The other features of $E_U$ can also be computed (deterministically) within this time bound. This completes the proof of the theorem. $\square$

**Corollary G.3.** *If all the polyhedra of $\mathcal{P}$ are unbounded in one of the two directions parallel to $\ell_0$, then the complexity of $\mathcal{T}_{\ell_0}(\mathcal{P})$ is $O(nk\beta_4(k))$, and there is a deterministic algorithm which computes $\mathcal{T}_{\ell_0}(\mathcal{P})$ in $O(nk(\log n + \alpha(k)\log k))$ time.*

*Proof.* Suppose that the direction in which the polyhedra are unbounded is the positive $z$-direction. In this case the functions $\sigma_P^+$ are all undefined, and the stabbing region is the region above the upper envelope $E_U$ of the functions $\sigma_P^-$, for $P \in \mathcal{P}$. The claim is then immediate from Theorem 4.2. $\square$

**Remarks:** It would be nice to extend Theorem 4.2 to the sandwich region between the two envelopes $E_U$ and $E_L$, as in (1). The difficulty in obtaining such an extension is that the overshadowing relation is not a necessary condition for two cross-sections $P(\theta_0), Q(\theta_0)$ to have two common tangents which are, say, lower tangents to $P(\theta_0)$ and upper tangents to $Q(\theta_0)$; see Figure 10 (left). Hence, the reduction to 2-dimensional sandwich regions, as in the proof of Theorem 4.2, does not carry over to the "mixed" case. The analysis in Section 3 is the way we have managed to overcome this problem, at the cost of a slight degradation in the bound. See also a discussion in Section 5.

## H   A Note on Geometric Permutations

In this section we apply the machinery developed in this paper to establish Theorem 4.3[9]:

**Theorem H.1.** *Let $\mathcal{P}$ be a collection of $k$ pairwise disjoint convex objects in $\mathbb{R}^3$, one of which is a line $\ell_0$. Then the number of geometric permutations induced by $\mathcal{P}$ is $O(k^3)$.*

*Proof.* For each convex set $C \in \mathcal{P} \setminus \{\ell_0\}$, let $h_C$ be a plane passing through $\ell_0$ and disjoint from $C$. The collection of the planes $h_C$ partitions $\mathbb{R}^3$ into (up to) $2(k-1)$ wedges, with the following property, which extends the discussion in the preceding section: For each of these wedges $W$, and for any directed line $\ell$, which is a transversal of $\mathcal{P}$ so that its forward ray $\ell^+$ from $\ell_0$ is contained in $W$, the set $\mathcal{P}^+$ of the objects of $\mathcal{P} \setminus \{\ell_0\}$ that are crossed by $\ell^+$ is the same for all such lines $\ell$ (it depends only on $W$). Clearly, this also holds for the complementary set $\mathcal{P}^-$ of objects crossed by the backward ray $\ell^-$ of $\ell$. Denote by $\mathcal{W}$ the set of these wedges.

Next, for each pair of objects $C, C'$ in $\mathcal{P} \setminus \{\ell_0\}$, choose an arbitrary plane $h_{C,C'}$ which separates $C$ and $C'$; for simplicity, we may assume that $h_{C,C'}$ is not parallel to $\ell_0$. Let $z_{C,C'}$ denote the point $\ell_0 \cap h_{C,C'}$. We may also assume that all the points $z_{C,C'}$ are distinct. These points partition $\ell_0$ into $\binom{k-1}{2} + 1$ intervals, and we denote by $\mathcal{I}$ the set of these intervals; see Figure 10 (right).

Each directed line $\ell$ which passes through $\ell_0$ can be labeled by the pair $(W, I) \in \mathcal{W} \times \mathcal{I}$, where $W$ is the wedge containing the forward ray of $\ell$, and $I$ is the interval containing the intercept of $\ell$ with $\ell_0$. We claim that, for each fixed pair $(W, I)$, all transversals of $\mathcal{P}$ labeled by $(W, I)$ generate the same geometric permutation. Indeed, let $\ell$ be such a transversal. As argued above, the sets $\mathcal{P}^+, \mathcal{P}^-$, consisting of those objects of $\mathcal{P} \setminus \{\ell_0\}$ crossed by the forward and backward rays $\ell^+, \ell^-$ of $\ell$, respectively, are fixed, and are independent of $\ell$. Moreover, we can sort, in a unique manner independent of $\ell$, the elements of $\mathcal{P}^+$ along $\ell^+$, as follows. For any pair of sets $C, C' \in \mathcal{P}^+$, consider the separating plane $h_{C,C'}$. By construction, $I$ lies fully on one side of $h_{C,C'}$, say the side containing $C$. But then $C$ must precede $C'$ along $\ell^+$, for otherwise $\ell^+$ would have to cross $h_{C,C'}$ twice, once from a point in $\ell_0$ (i.e., in $I$) to a point in $C'$ and then back to a point in $C$. See Figure 10 (right). Hence the order of any two elements of $\mathcal{P}^+$ along $\ell^+$ is fixed for all lines $\ell$ labeled by $(W, I)$, so the suffix of the permutation induced by $\ell$, from $\ell_0$ on, is fixed too. A symmetric argument shows that the prefix of the permutation, up to $\ell_0$, is also fixed, so the permutation itself is fixed for all lines labeled by $(W, I)$. Since the number of labels is $O(k^3)$, the theorem follows. □

**Remarks:** (a) Clearly, the proof shows that the number of geometric permutations is at most $O(Hk)$, where $H$ is the smallest cardinality of a set of planes which separate $\mathcal{P}$ (that is, each pair of objects of $\mathcal{P}$ is separated by some plane in the set). Unfortunately, there are constructions of sets of $k$ pairwise disjoint polyhedra for which $H = \Theta(k^2)$. For example, consider the hyperbolic paraboloid $z = xy$, and draw on it two sets, $L_1, L_2$, of generating lines, where $L_1$ consists of the lines $x = i$, $z = iy$, for $i = 1, \ldots, k/2$, and $L_2$ consists of the lines $y = j$, $z = jx$, for $j = 1, \ldots, k/2$. Now keep $L_1$ intact, and shift $L_2$ upwards by some $\varepsilon > 0$. It is then easy to check that, if $\varepsilon$ is chosen sufficiently small, for each pair of lines $\ell_1 \in L_1$, $\ell_2 \in L_2$, any plane that separates $\ell_1$ and $\ell_2$ must intersect all the other lines in $L_1 \cup L_2$, so $H \geq k^2/4$.

---

[9]The proof uses a technique reminiscent of, and borrowing some ideas from, the analysis in [5].

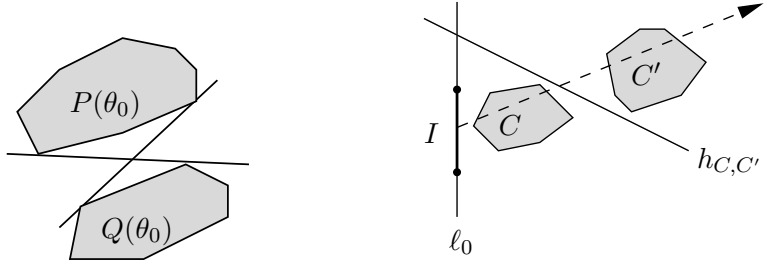

Figure 10: The overshadowing relation is not necessary for the cross-sections $P(\theta), Q(\theta)$ to have two common tangents (left). Geometric permutations through $\ell_0$: the interval $I$ lies on one side of $h_{C,C'}$ (right).

(b) Note that the proof of Theorem 4.3 fails if we replace $\ell_0$ by a *line segment*, because it may then be impossible to separate the sets of $\mathcal{P}$ from $\ell_0$ by planes passing through $\ell_0$.